\documentclass[a4paper,11pt]{article}
\pdfoutput=1 

\usepackage{jheppub} 

\usepackage[T1]{fontenc} 

\title{\boldmath Superadiabatic basis in cosmological particle production: application to preheating}

\author[a]{Yusuke Yamada}

\affiliation[a]{Research Center for the Early Universe (RESCEU), Graduate School of Science,\\ The University of Tokyo, Hongo 7-3-1,
Bunkyo-ku, Tokyo 113-0033, Japan}
\emailAdd{yamada@resceu.s.u-tokyo.ac.jp}

\preprint{RESCEU-12/21}

\abstract{We discuss the adiabatic basis dependence of particle number in time-dependent backgrounds. In particular, we focus on preheating after inflation, and show that, for the optimal basis, the time dependence of the produced particle number can be well approximated by a simple connection formula, which can be obtained by analysing Stokes phenomenon in given backgrounds. As we show explicitly, the simple connection formula can describe various parameter regions such as narrow and broad resonance regime in a unified manner. }
\begin{document}
\maketitle
\flushbottom

\section{Introduction}
Physics in the early Universe would be effectively described by quantum field theory coupled to classical fields, such as the expanding Universe background. In particular, during or after inflation, classical inflaton field or possibly multiple fields also appear as such background fields. In the presence of time-dependent background, quantum fields coupled to them show nontrivial behavior that is absent for quantum fields in time-independent backgrounds. One of the most nontrivial properties is particle production from ``vacuum'' such as Sauter-Schwinger effect~\cite{Sauter:1931zz,Schwinger:1951nm} and Hawking radiation~\cite{Hawking:1974sw}. Graviational particle production was also found in early works of Parker~\cite{Parker:1969au,Parker:1971pt}. Understanding of such phenomenon would be interesting from theoretical as well as phenomenological viewpoints. 

The particle production in time-dependent backgrounds is well understood as {\it Stokes phenomenon}, which is related to asymptotic behavior of functions. Such a viewpoint leads to a systematic approach to evaluate particle production in various models~\cite{Dumlu:2010ua,Dumlu:2010vv,Li:2019ves,Dumlu:2020wvd,Hashiba:2020rsi,Enomoto:2020xlf,Taya:2020dco,Hashiba:2021npn,Sou:2021juh}. In time-dependent backgrounds, we define ``particles'' or ``vacuum'' by specifying boundary conditions on mode functions, especially by defining positive and negative frequency modes, which are not uniquely determined unlike that in time-independent backgrounds. Due to the time-dependence of the effective frequency of each mode function, there is no unique way to separate positive and negative frequency modes. Nevertheless, we are able to define positive and negative frequency modes e.g. by the Wenzel-Kramers-Brillouin (WKB) method. Thus defined two modes in a particular time region will ``mix'' with each other by Stokes phenomenon, which is understood as ``particle production''.\footnote{ For review of Stokes phenomenon, see e.g.~\cite{Froman1} for mathematical details of Stokes phenomenon or \cite{Hashiba:2021npn}, which also discusses applications to gravitational particle production.} The Stokes phenomenon takes place when we cross the so-called Stokes lines, which separate the region where positive and negative frequency modes can be definite.\footnote{This point would be more clear in the context of exact WKB analysis. For exact WKB solutions, the Stokes line is the region where Borel sum is impossible. Therefore, one needs to consider analytic continuation of mode functions when crossing Stokes lines on the complex time plane. See e.g.~\cite{Taya:2020dco} for more detailed explanations of exact WKB analysis. In this work, we will use WKB or more generally the phase integral method~\cite{Froman1}.} The connection formula of positive and negative frequency modes at Stokes line crossing can be found in a very systematic way. 

The notion of Stokes phenomenon can be understood from the asymptotic series. In cosmological context, it is also known as the adiabatic expansion, which is applied for regularization of ultraviolet divergences in time-dependent backgrounds~\cite{Zeldovich:1971mw,Parker:1974qw}. In particular, in the analysis of asymptotics, Dingle found a smooth connection formula at Stokes line crossing~\cite{Dingle} and Berry derived it using partial Borel resummation~\cite{Barry:1989zz}. Berry also found that the smooth connection formula describes the behavior of solutions with the optimal adiabatic basis, called {\it superadiabatic basis}~\cite{Dabrowski:2014ica,Dabrowski:2016tsx}. It was explicitly shown that indeed the solutions with the superadiabatic basis are well described by the Dingle-Berry formula~\cite{Dabrowski:2014ica,Dabrowski:2016tsx} in the context of Sauter-Schwinger effects. 

In this work, we discuss the (super)adiabatic basis in preheating. Preheating after inflation~\cite{Kofman:1994rk,Kofman:1997yn} leads to explosive particle production from ``vacuum''.\footnote{See also~\cite{Amin:2014eta} for a recent review on preheating.} We first consider parametric resonance of a spectator scalar field coupled to an oscillating scalar field~\cite{Dolgov:1989us,Traschen:1990sw,Kofman:1994rk,Shtanov:1994ce,Yoshimura:1995gc,Kofman:1997yn}, which captures the property of preheating. We emphasize that the purpose of this paper is not to understand preheating itself. The dynamics of preheating was comprehensively studied in the pioneering work by Kofman, Linde and Starobinsky~\cite{Kofman:1997yn}. Our interest is to give a different description of particle production in preheating from Stokes phenomenon viewpoints, which give us more systematic ways to analyse the particle production dynamics in various background field configurations. It is also important to explicitly show that the particle number defined by the superadiabatic basis can be well approximated by the Dingle-Berry's error function formula. We would like to emphasize that our approach does not rely on the property of Mathieu equation, which plays important roles in understanding preheating.

One of our important results is that our approach gives a unified way to describe the narrow and broad parametric resonance. The former can be understood from perturbative effect with Bose enhancement, whereas the latter is non-perturbative in the coupling constant. Despite such a different parameter dependence, our analysis based on Stokes phenomenon captures both regime by just changing the parameters. This implies that Stokes phenomenon can describe both perturbative and non-perturbative particle production.

The rest of this paper is organized as follows. In Sec.~\ref{adrev}, we would like to review adiabatic expansion discussed in~\cite{Dabrowski:2016tsx}, which seems less known in the context of cosmology. We also briefly discuss the relation between particle production and Stokes phenomenon, which is also important for determining the optimal adiabatic basis. In Sec.~\ref{comments}, we briefly discuss the relevance of the superadiabatic basis to physical quantities. In Sec.~\ref{adpart}, we apply the superadiabatic basis and analysis of Stokes phenomenon to preheating and its toy models. These examples clarify the difference of the adiabatic basis and that the optimal adiabatic basis can be well approximated by the formula found by Dingle and Berry, or its extension shown in appendix~\ref{univform}. Finally, we conclude in Sec.~\ref{concl}. 

Throughout this work, we use the natural unit $c=1,\hbar=1$. The metric sign convention is $(-+++)$.

\section{Adiabatic basis: review}\label{adrev}
\subsection{``standard'' method}
We review a commonly used definition of adiabatic vacuum, and will discuss its ambiguity in the next subsection.
In this section, we will consider a scalar field with a time-dependent mass,
\begin{eqnarray}
\mathcal{L}=-\frac12 (\partial\chi)^2-\frac12 m^2(t)\chi^2,
\end{eqnarray}
and we Fourier expand the scalar field as
\begin{eqnarray}
    \hat{\chi}(t,{\bf x})=\int \frac{d^3k}{(2\pi)^{\frac32}}e^{{\rm i}{\bf k}\cdot{\bf x}}\left[\hat{a}_{\bf k}v_k(t)+\hat{a}_{-{\bf k}}^\dagger v^*_{k}(t)\right],
\end{eqnarray}
where $v_{k}(t)$ is a mode function, $\hat{a}_{\bf k}$ and  $\hat{a}_{\bf k}^\dagger$ are creation and annihilation operator, respectively, which satisfy $[a_{\bf k},a^\dagger_{{\bf k}'}]=\delta^3({\bf k}-{\bf k}')$. The mode function $v_k(t)$ should follow the normalization condition,
\begin{equation}
    v_k\dot{v}_k^*-\dot{v}_kv_k^*=-{\rm i},\label{norm}
\end{equation}
at any time $t$. The mode equation is given by
\begin{equation}
    \ddot{v}_k+\omega_{k}^2v_k=0,\label{mode}
\end{equation}
where $\omega_k^2=m^2(t)+k^2$. 

The WKB method is commonly used in order to estimate the particle production rate, where we choose the positive and the negative basis function to be 
\begin{equation}
    f_k^{\pm}=\frac{1}{\sqrt{2\omega_k}}e^{\mp {\rm i}\int_{t_0}^t\omega_k(t')dt'},
\end{equation}
where $t_0$ is a reference time often taken to be $t_0\to-\infty$. With these basis functions, we can formally find the mode function to be
\begin{equation}
    v_k(t)=\alpha_k(t)f_k^++\beta_k(t)f_k^-.\label{wkb}
\end{equation}
Here, the auxiliary functions $\alpha_k(t)$ and $\beta_k(t)$ satisfy
\begin{eqnarray}
   \dot{\alpha}_k(t)&=&\frac{\dot{\omega}_k}{2\omega_k}e^{2{\rm i}\int_{t_0}^t\omega_kdt'}\beta_k(t),\label{aeq}\\
  \dot{\beta}_k(t)&=&\frac{\dot{\omega}_k}{2\omega_k}e^{-2{\rm i}\int_{t_0}^t\omega_kdt'}\alpha_k(t),\label{beq}
\end{eqnarray}
so that the mode function \eqref{wkb} be the solution of the mode equation~\eqref{mode}. Assuming $m^2(t)\to \text{const}$ as $t\to \pm\infty$, we are able to define ``particles'' in both the future and past infinity.\footnote{We will consider the backgrounds that are not asymptotically static. In this sense, it would be better to call thus defined ``vacuum'' as instantaneous adiabatic vacuum.}

Within this formalism, the specification of the vacuum corresponds to the choice of the boundary conditions on $\alpha_k$ and $\beta_k$. The past adiabatic vacuum condition is 
\begin{equation}
    \alpha_k(-\infty)=1,\beta_k(-\infty)=0.\label{ic}
\end{equation}
From the future observer, the number density of the particle produced by the time variation of the mass is given by $n_k=|\beta_k(\infty)|^2$. In most cosmological models, it is not always possible to assume $m(t)^2\to {\rm const}$ as $t\to -\infty$. In such a case, we need to impose conditions like \eqref{ic} at some reference time $t_0$. As long as the total number density $n(t)=\int d^3k |\beta_k(t)|^2$ is finite, thus defined instantaneous vacuum can have overlap with the vacuum at different time $t$. In other words, if $n(t)$ were not finite, such a vacuum state is not well defined~\cite{Parker:1969au,Fulling:1979ac}.

When we only discuss the particle number at the future and past infinity where the particle number is definite, the commonly used WKB method is sufficient to estimate the produced particle number. However, if we are interested in the particle number at an {\it intermediate} time, we should be careful: One may think that $n_k(t)=|\beta_k(t)|^2$ derived by solving \eqref{aeq} and \eqref{beq} describes the particle number at a given time $t$. However, the ``particle number'' at intermediate time depends on the choice of the basis function $f_k^\pm$. As we will discuss below, the choice of the basis function is not unique, and in turn, the auxiliary functions $\alpha_k(t)$ and $\beta_k(t)$ are not uniquely determined.

\subsection{Superadiabatic basis}
Following~\cite{Dabrowski:2014ica,Dabrowski:2016tsx}, we review more general definition of adiabatic vacuum. We use the following ansatz of the mode function 
\begin{equation}
    v_k=\frac{1}{\sqrt{2W_k}}e^{-{\rm i}\int_{t_0}^tW_kdt'},
\end{equation}
where the function $W_k(t)$ is an unspecified function. Substituting this ansatz to \eqref{mode}, we find the condition on $W_k(t)$
\begin{equation}
W_k^2(t)=\omega_k^2(t)-\left[\frac{\ddot{W}_k}{2W_k}-\frac{3}{4}\left(\frac{\dot{W}_k}{W_k}\right)^2\right].
\end{equation}
Finding the solution $W_k(t)$ is equivalent to solve the mode equation {\it exactly}. However, it is impossible in general. Therefore, we will instead consider approximations: Let us assume adiabatic behavior of $\omega_k$ namely $|\dot{\omega}_k/\omega_k^2|\ll1$, which may be violated at some region. Assuming adiabaticity, we are able to solve the equation by iteration as
\begin{equation}
   ( W_k^{(j+1)})^2=\omega_k^2-\left[\frac{\ddot{W}_k}{2W_k}-\frac{3}{4}\left(\frac{\dot{W}_k}{W_k}\right)^2\right]\Biggr|_{W_k=W_k^{(j)}},
\end{equation}
where $j=0,1,2,\cdots$ with $W_k^{(0)}=\omega_k$ as the lowest solution. At the $(j+1)$-th order, one has to take the terms containing at most $2(j+1)$ time derivatives. For example, the first order is given by
\begin{equation}
    ( W_k^{(1)})^2=\omega_k^2-\left[\frac{\ddot{\omega}_k}{2\omega_k}-\frac{3}{4}\left(\frac{\dot{\omega}_k}{\omega_k}\right)^2\right].
\end{equation}
This is known as adiabatic expansion, which is often used to regularize the UV divergence in time dependent background such as expanding Universe~\cite{Zeldovich:1971mw,Parker:1974qw}. This can be thought of the generalization of the WKB approximation, which corresponds to the 0-th order in the adiabatic expansion. It is known that the adiabatic expansion is asymptotic series, namely the series does not converge. Therefore, we need to truncate the series at the optimal order. The optimal order can be estimated by the phase integral along the turning points as shown in appendix~\ref{univform}. For derivation of the estimate, we refer \cite{Barry:1989zz,Li:2019ves,Sou:2021juh}.

For any adiabatic order, we are able to construct the solution of the mode equation with the help of auxiliary functions as we showed in the previous section. Let us take the $j$-th order, and we formally write the mode function as
\begin{equation}
v_k(t)=\alpha_k(t)f^{+(j)}_k+\beta_k(t)f_k^{-(j)},\label{mdj}
\end{equation}
where
\begin{equation}
    f_k^{\pm (j)}\equiv \frac{1}{\sqrt{2W_k^{(j)}}}\exp\left[\mp{\rm i}\int_{t_0}^tW_k^{(j)}dt'\right]
\end{equation}
are the $j$-th order basis functions and $\alpha_k(t)$ and $\beta_k(t)$ are unspecified auxiliary functions, which play the role of the Bogoliubov coefficients. We require the first time-derivative of $v_k$ to be
\begin{equation}
    \dot{v}_k(t)=(-{\rm i}W_k^{(j)}+V_k)\alpha_kf_k^{+(j)}+({\rm i}W_k^{(j)}+V_k)\beta_kf_k^{-(j)},\label{1stder}
\end{equation}
where $V_k(t)$ is a real function of time. We can easily check that the normalization condition $v_k\dot{\bar{v}}_k-\dot{v}_k\bar{v}_k={\rm i}$ is satisfied if $|\alpha_k(t)|^2-|\beta_k(t)|^2=1$ independently of the real function $V_k(t)$. Therefore, $V_k(t)$ is another ambiguity of the definition of the basis function. The ``natural choice'' of this degree of freedom is proposed in~\cite{Dabrowski:2016tsx} as $V_k=-\frac{\dot{W}_k^{(j)}}{2W_k^{(j)}}$.

In order for $\alpha_k(t)$ and $\beta_k(t)$ to satisfy \eqref{1stder} and the mode equation, the evolution equations of $\alpha_k$ and $\beta_k$ should be
\begin{equation}
    \left(\begin{array}{c}\dot{\alpha}_k(t)\\ \dot{\beta}_k(t)\end{array}\right)=\left(\begin{array}{cc}\delta_k&(\Delta_k+\delta_k)e^{2{\rm i}\int_{t_0}^tW_k^{(j)}dt'}\\
  (\Delta_k-\delta_k)e^{-2{\rm i}\int_{t_0}^tW_k^{(j)}dt'} &-\delta_k \end{array}\right)\left(\begin{array}{c}\alpha_k(t)\\ \beta_k(t)\end{array}\right),\label{Beq}
\end{equation}
where 
\begin{eqnarray}
    \delta_k&\equiv& \frac{1}{2{\rm i}W_k^{(j)}}(\omega_k^2-(W^{(j)}_k)^2+\dot{V}_k+V_k^2),\\
    \Delta_k&\equiv& \frac{\dot{W}_k^{(j)}}{2W_k^{(j)}}+V_k.
\end{eqnarray}
Here, we have left $V_k$ in a general form, and the ``standard'' WKB formula discussed in the previous section corresponds to the choice $V_k(t)=0$ with $W_k^{(0)}=\omega_k(t)$. The natural choice $V_k=-\frac{\dot{W}_k^{(j)}}{2W_k^{(j)}}$ simplifies the expression because $\Delta_k=0$. The simplified differential equations for the Bogoliubov coefficients are
\begin{equation}
    \left(\begin{array}{c}\dot{\alpha}_k(t)\\ \dot{\beta}_k(t)\end{array}\right)=\delta_k\left(\begin{array}{cc}1&e^{2{\rm i}\int_{t_0}^tW_k^{(j)}dt'}\\
  -e^{-2{\rm i}\int_{t_0}^tW_k^{(j)}dt'} &-1 \end{array}\right)\left(\begin{array}{c}\alpha_k(t)\\ \beta_k(t)\end{array}\right),
\end{equation}
and one can easily confirm that the condition $|\alpha_k(t)|^2-|\beta_k(t)|^2=1$ is satisfied at any time~$t$. 

As explicitly seen from the discussion so far, the definition of the basis function or equivalently $(\alpha_k(t),\beta_k(t))$ is indeed ambiguous. We may interpret this ambiguity as follows. The choice of basis functions is the definition of energy quantum, which accordingly defines a ``particle''. Since the time translation invariance is broken by the background field, we have no unique choice of the energy quanta defining particles. Nevertheless, the optimal choice of the ``energy'' quantum would give us the optimal definition of a ``particle''. We may summarize the construction so far as follows. First, we need to define the energy quantum, namely, need to choose the adiabatic order of $W_k^{(j)}$. This is yet insufficient to define a particle. The ambiguity comes from the first order derivative $\dot\chi$. We have a real function degree of freedom $V_k(t)$, which determines the first order differential equation of the auxiliary functions $\alpha_k(t)$ and $\beta_k(t)$. Once we fix the real function $V_k(t)$, an ``adiabatic particle'' is defined. The adiabatic vacuum state corresponds to the boundary conditions for $\alpha_k,\beta_k$.

We also note that in~\cite{Dabrowski:2016tsx} it is shown that there are several approaches to obtain adiabatic particle spectra, Klein-Gordon equation, Ermakov-Milne equation, Riccati equation, and spectral function approach. The authors of~\cite{Dabrowski:2016tsx} show the equivalence between them. In the following discussion, we will take the Klein-Gordon equation approach, which is more commonly used to discuss e.g. particle production.\footnote{For numerical simulations, Ermakov-Milne equation seems useful, and we review it in appendix~\ref{EMeq}.} Nevertheless, the equivalence shown in \cite{Dabrowski:2016tsx} ensures any approach gives the same result.
\begin{figure}[htbp]
\centering
\includegraphics[width=.5\textwidth]{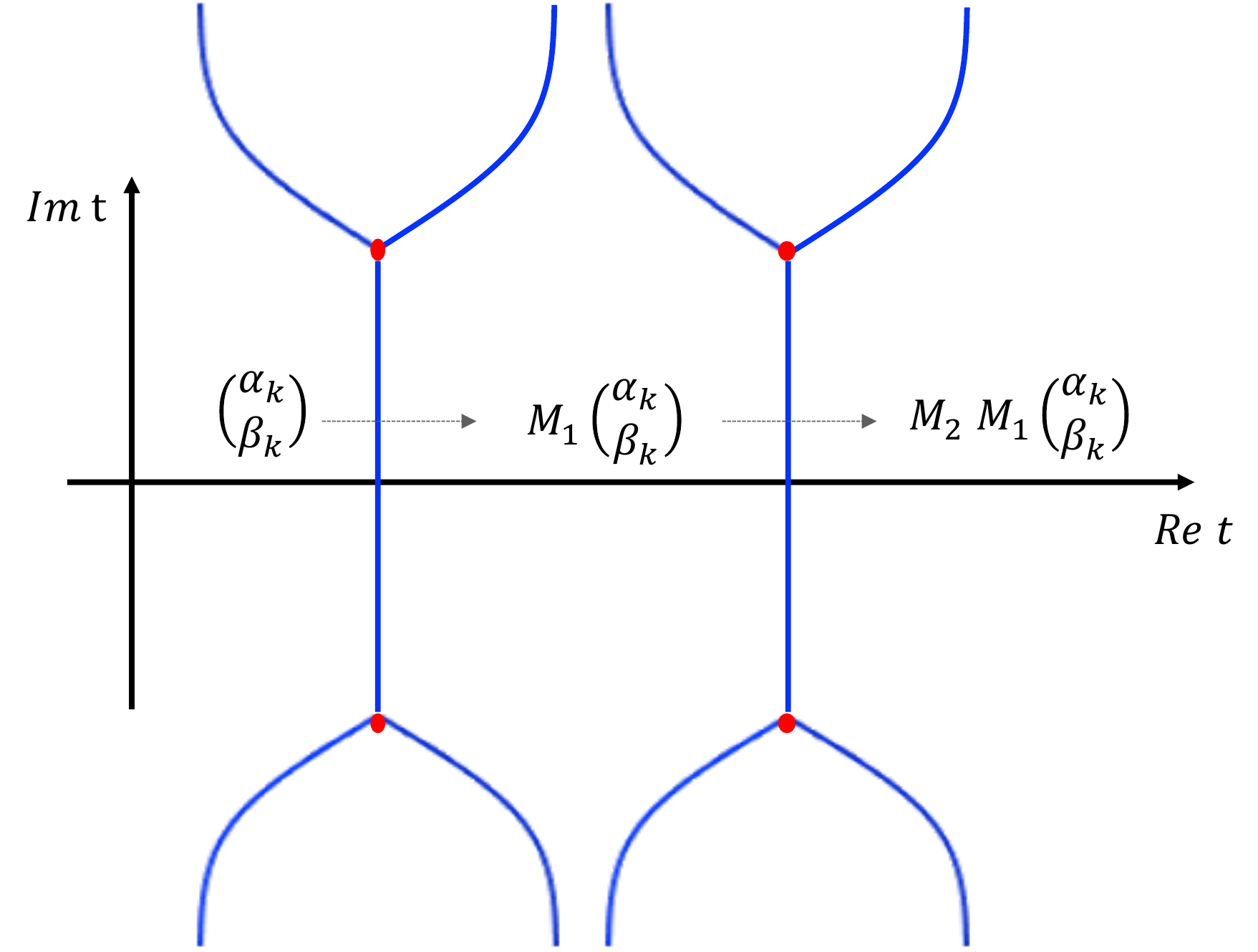}
\caption{Schematic picture of Stokes phenomenon. Turning points are denoted by red dots, and the Stokes lines (blue lines) emanate from each turning point. Stokes lines cross the real axis, which corresponds to particle production ``events''. When crossing Stokes lines, $(\alpha_k,\beta_k)$ change, which can be effectively described by multiplications of matrix $M_{i}$.}\label{fig;stokes pheno}
\end{figure}
\subsection{Stokes phenomenon and particle production}
In this section, we briefly review the Stokes phenomenon and its relation to particle production. We refer \cite{Froman1} for more details of Stokes phenomenon and WKB/phase integral methods (see also \cite{Hashiba:2021npn}). Instead of a general review of Stokes phenomena, we will explain how to evaluate the connection formula associated with Stokes phenomenon in the next section within simple examples. The explanation here is rather abstract.

The particle production caused by time-dependent backgrounds can be understood from asymptotic behavior of mode functions. In order to understand the asymptotic behavior, the WKB method or more generally the phase integral method is quite useful. For a given mode equation, the analytic property of solutions is characterized by $\omega_k^2(t)$. The behavior of the solution on complex $t$-plane significantly changes around turning points $t_c$ at which the frequency vanishes $\omega_k^2(t_c)=0$. From turning points, {\it Stokes lines} emanate, on which ${\rm i}\int^t \omega_k dt$ is real. On the stokes line, the mode function $e^{\pm{\rm i}\int\omega_k dt}$ increases or decreases significantly. Such lines ends up at poles of $\omega_k$, different turning points or infinity. Thus, Stokes lines separate the complex plane into several pieces. When we consider analytic continuation of basis functions from one region to another separated by Stokes lines, the analytic continuation leads to non-trivial mixing between positive and negative frequency modes, which may give rise to nonvanishing $\beta_k$ even if we take $\beta_k=0$ namely an adiabatic vacuum condition at an initial time. See Fig.~\ref{fig;stokes pheno}. The nontrivial mixing between positive and negative frequency basis functions is what we call Stokes phenomenon.\footnote{In Fig.~\ref{fig;stokes pheno}, we describe the Stokes phenomenon by multiplication of matrices on $\alpha_k,\beta_k$. This is equivalent to mixing of positive and negative frequency modes.} Even if turning points do not appear on real $t$-axis, Stokes lines cross real $t$-axis and therefore Stokes phenomenon, or equivalently, ``particle production'' necessarily takes place. 

The adiabatic expansion is in general an asymptotic series, namely the series does not converge. Therefore, it is not mathematically well-defined. In order to have a meaningful solution, we may consider Borel transformation of adiabatic series, which gives well-defined convergent solutions. Such an analysis is known as exact WKB analysis. (See \cite{Taya:2020dco} and references therein.) From exact WKB analysis viewpoints, the Stokes line corresponds to the place where the series are not Borel summable, and therefore the exact WKB solution is discontinuous on Stokes lines. Therefore, one needs to consider analytic continuations of solutions when crossing Stokes lines. The analytic continuation gives non-trivial mixing of positive and negative frequency solutions as mentioned above.

Berry applied partial Borel resummation technique to later terms of the adiabatic expansion and showed a smooth connection formula of $\beta_k$ around the Stokes line crossing~\cite{Barry:1989zz}, which turn out to be a good analytical formula describing the ``time-dependent particle numbers''~\cite{Dabrowski:2014ica,Dabrowski:2016tsx}. We summarize the formula in Appendix~\ref{univform} without derivations, and refer to Berry's original paper~\cite{Barry:1989zz} for the derivation. (See also recent papers~\cite{Li:2019ves,Sou:2021juh}.)

\section{Physical relevance of adiabatic basis choices}\label{comments}
In this section, we briefly discuss how the choice of adiabatic basis affects {\it physical quantities}. If we are only interested in the final value of physical quantities, we are able to evaluate it via $(\alpha_k(\infty),\beta_k(\infty))$, which is independent of the basis choice. On the other hand, in order to know the time evolution of physical quantities such as energy density at intermediate time, we need to solve the evolution equation of $(\alpha_k(t),\beta_k(t))$. The important point is that for the superadiabatic basis, $\alpha_k(t)$ and $\beta_k(t)$, we are able to apply Dingle-Berry's approximate formula. It would mean that we are able to obtain an approximation for physical quantities written in terms of basis functions and $(\alpha_k(t),\beta_k(t))$. Besides that, ``optimal'' particle picture would be useful if we are interested in ``scattering events'' on time-dependent backgrounds.\footnote{Strictly speaking, it would be difficult to discuss particle scattering events in time-dependent backgrounds as the case of Minkowski spacetime, since asymptotic freeness and completeness are absent in time-dependent backgrounds. Nevertheless, with optimal adiabatic particle, we would be able to have e.g. correct quantum kinetic equations or Boltzmann equations in a given time dependent backgrounds.}

Let us show the dependence of the adiabatic basis on physical quantities. For concreteness, hereafter we will focus on the following model:
\begin{equation}
S=\int d^4x\sqrt{-g}\left[-\frac12(\partial\phi)-V(\phi)-\frac12 (\partial\tilde{\chi})^2-\left(\frac{m^2}{2}+\frac{\lambda}{2} h(\phi)\right)\tilde{\chi}^2\right],\label{action}
\end{equation}
where $\phi$ and $\tilde{\chi}$ denote real scalar fields, $V(\phi)$, $h(\phi)$ are functions of $\phi$, $m$ is the mass of $\chi$, and $\lambda$ and $g$ denote real coupling constants. In the following discussion, we will regard $\phi$ and metric as the homogeneous background classical field $\phi=\phi(t), \ g_{\mu\nu}=g_{\mu \nu}(t)$ and ignore its fluctuations around the background. 
We take the background geometry to be flat Friedman-Robertson-Walker (FRW) spacetime $ds^2=-dt^2+a^2(t)d{\bf x}^2$, where $a(t)$ is the scale factor. On this background, the canonical quantum field $\hat{{\chi}}\equiv a^{3/2}\hat{\tilde{\chi}} $ can be expanded as
\begin{equation}
   \hat{\chi}=\int \frac{d^3k}{(2\pi)^{3/2}}e^{{\rm i}{\bf k}\cdot{\bf x}}[\hat{a}_{\bf k}v_k(t)+\hat{a}_{-{\bf k}}^\dagger v^*_{k}(t)],
\end{equation}
where the mode function $v_k$ satisfies
\begin{equation}
    \ddot{v}_k+\omega_k^2v_k=0,\label{modeeq}
\end{equation}
where 
\begin{equation}
    \omega_k^2=\frac{k^2}{a^2}+m^2+\lambda h(\phi)+\frac{9H^2}{4}-\frac{3\dot{H}}{2},\label{genomg}
\end{equation}
and $H=\dot{a}/a$. We normalize mode functions as \eqref{norm}. We decompose the mode functions as \eqref{mdj}, namely, we take $j$-th order adiabatic expansion.

Since we consider a free field, all the physical quantity are simply given by two point functions, or equivalently the products of mode functions. The energy density of $\chi$ is given by
\begin{align}
    \langle\rho_\chi\rangle=&\left\langle \frac{1}{2}\dot{\hat{\tilde{\chi}}}^2+\frac{1}{2a^2}(\partial_i\hat{\tilde{\chi}})^2+\frac{1}{2}(m^2+\lambda h(\phi))\hat{\tilde{\chi}}^2\right\rangle\nonumber\\
    =&\left\langle \frac{1}{2a^3}\dot{\hat\chi}^2+\frac{1}{2a^5}(\partial_i{\hat{\chi}})^2+\frac{1}{2a^3}\left(m^2+\lambda h(\phi)+\frac{9H^2}{4}-\frac{3\dot{H}}{2}\right)\hat{\chi}^2\right\rangle.
\end{align}
These expectation values can be expressed by the background quantities and $v_k$. Since we are able to reexpress $v_k$ by the basis function and $(\alpha_k(t),\beta_k(t))$, the energy density can be evaluated as
\begin{equation}
    \langle\rho_\chi\rangle=\frac{1}{2a^3}\int \frac{d^3k}{(2\pi)^3}\biggl[\frac{1}{2W_k^{(j)}}(|Q_k|^2+\omega_k^2)+\frac{|\beta_k|^2}{W_k^{(j)}}(|Q_k|^2+\omega_k^2)+\left(\alpha_k\bar{\beta}_k(f^{+(j)}_k)^2(Q^2+\omega_k^2)+{\rm h.c.}\right)\biggr], \label{energy}
\end{equation}
where $Q_k\equiv -{\rm i}W_k^{(j)}+V_k$ and we have used the normalization condition $|\alpha_k|^2-|\beta_k|^2=1$. 

Let us understand the meaning of terms in \eqref{energy}. The first term of \eqref{energy} has no dependence on $\alpha_k$ or $\beta_k$, and we can interpret it as vacuum contribution to the energy density. Indeed, if we take the WKB choice $W_k^{(j)}=\omega_k,\ V_k=0$ and static limit $\omega_k\to {\rm const}, a(t)=1$, we find the vacuum energy $\rho\sim (2\pi)^{-3} \int d^3k \omega_k/2$. This contribution needs to be renormalized by appropriate methods as usual. One of the standard methods would be adiabatic regularization~\cite{Zeldovich:1971mw,Parker:1974qw}, where we subtract the divergent quantities evaluated with the local adiabatic vacuum state with some order of the adiabatic expansion. For example, in order to renormalize energy density $\rho$, one needs to subtract energy density calculated with the 4-th order adiabatic expansion with $\alpha_k=1,\ \beta_k=0$~\cite{Parker:1974qw}. Physically speaking, this procedure means that we subtract locally defined vacuum contribution.\footnote{For more details of the adiabatic regularization, see e.g.~\cite{Parker:2009uva}. We also note that commonly used definition of the adiabatic expansion order $j'$ is different from what we refer as the adiabatic order $j$. Their relation is $j'=2j$.} Since $\beta_k(t)$ is determined by Stokes phenomenon, which is related to the global structure of $\omega_k(t)$, local counterterms cannot remove the terms containing $\beta_k$, whereas the vacuum energy renormalization is possible since it is the subtraction of ``local'' contribution of microscopic modes.\footnote{As we mentioned above, for the adiabatic regularization, we calculate the vacuum contribution by evaluating physical quantities with $\alpha_k=1,\beta_k=0$. Therefore, the renormalization procedure does not change terms including $\beta_k(t)$. This is the reason why we will focus on the terms in \eqref{rhonp}.} The renormalized vacuum contribution depends on the renormalization condition we impose, and therefore we will not discuss the vacuum contribution in the following discussion. 

Next, we focus on the rest of \eqref{energy}, which would be meaningful almost independently of the renormalization schemes since $\beta_k$ is related to the particle production. We call this contribution as $\langle \rho_\chi\rangle_{\rm np}$, namely
\begin{eqnarray}
 \langle\rho_\chi\rangle_{\rm np}=\frac{1}{2a^3}\int \frac{d^3k}{(2\pi)^3}\biggl[
   \frac{|\beta_k|^2}{W_k^{(j)}}(|Q_k|^2+\omega_k^2)+\left(\alpha_k\bar{\beta}_k(f^{+(j)}_k)^2(Q_k^2+\omega_k^2)+{\rm h.c.}\right)\biggr].\label{rhonp}
\end{eqnarray}
This contribution contains non-perturbative effects in coupling constants. In order to understand the physical meaning of $\langle \rho_\chi\rangle_{\rm np}$, let us take the WKB choice $W_k^{(j)}=\omega_k(t)$ and $V_k=0$ and we find the oscillatory part proportional to $(f^{\pm(j)}_k)^2$ since $Q^2+\omega_k^2=0$ in this case. Then, we find $\langle\rho_\chi\rangle_{\rm np}= (2\pi a)^{-3}\int d^3k \omega_k n_k$, where $n_k=|\beta_k|^2$ is number density of a $k$-mode particle. This is nothing but the energy density of produced particles. From this fact, we may interpret $\langle\rho\rangle_{\rm np}$ as the generalized energy density induced by particle production. 

Let us briefly discuss the UV property of $\langle \rho_\chi\rangle_{\rm np}$. As will be shown in concrete examples, the typical value of $|\beta_k|$ decays exponentially for large $k$ modes if we correctly evaluate its value by taking Stokes phenomenon into account. We may understand the absence of the UV divergences as a consequence of the fact that Stokes phenomenon is not a local but a global property of $\omega_k(t)$. We may also expect that it cannot be divergent since it is non-perturbative effects, and as a result, the value of $\beta_k$ is not an analytic function of coupling constants. We would not be able to renormalize such a quantity by shifts of coupling constants if the contribution gave divergent contributions. We will clarify this point in the concrete examples shown later. Nevertheless, we do not mean local UV divergences are irrelevant to $\beta_k$. For instance, the vacuum energy contribution in the first term of \eqref{energy} depends on the external field $\phi$ with divergent coefficients, which would contribute to the equation of motion of $\phi$. Then, we need counter-terms in Lagrangian to remove the divergent terms as usual. After proper renormalization procedures, we expect finite quantum corrections to $\phi$'s equation of motion. Such corrections modify the dynamics of $\phi$ and accordingly, the time dependence of $\omega_k(t)$ of $\chi$ would be indirectly affected. Furthermore, if we consider interacting quantum field theory, renormalization of coupling constants and field strength would be necessary.\footnote{In time-dependent backgrounds, perturbation theory may fail due to the secular growth of effective couplings (see e.g.~\cite{Krotov:2010ma,Trunin:2018egi,Akhmedov:2021rhq}). Dynamical renormalization group technique may relax or remove the secular growth, which is applied to scalar field theory in de Sitter background (see e.g. \cite{Burgess:2010dd, Green:2020txs} and references therein). Or one needs to use 2PI effective action rather than the standard 1PI effective action, see e.g.~\cite{Berges:2004yj} for a review. We will not discuss these issues in this work.} As a result, again the structure of $\omega_k(t)$ would be corrected by quantum mechanical effects. Even in such a case, we expect that $\beta_k$ would decay exponentially for high momentum modes because of global nature of Stokes phenomenon. Related issues about renormalization are discussed e.g. in~\cite{Affleck:1981bma} in the context of the Schwinger effect.

We emphasize that the physical quantities should be independent of the basis choices as long as we fix the boundary conditions on $v_k$, despite the appearance of the basis dependent quantities in \eqref{energy}. This is because it is originally given by $v_k$. As obvious from construction, the basis choices are essentially how we decompose the mode function $v_k$ into the basis function and the auxiliary functions $(\alpha_k(t),\beta_k(t))$. The combination of them is always the solution of the mode equation. In this sense, the basis dependence on the physical quantities should disappear.\footnote{The separation of ``vacuum'' contribution and the other ``physical'' one depends on the adiabatic order. More precisely, the ``vacuum'' contribution depends on our choice of adiabatic basis. This is how the adiabatic regularization works for renormalization of divergent quantities. But, the regularized ``vaccuum'' part would also depend on the renormalization conditions. }

If physical quantities is independent of adiabatic basis we choose, what is the advantage of the superadiabatic basis? As we mentioned, for the superadiabatic basis, the time evolution of $(\alpha_k(t),\beta_k(t))$ can be approximated by the Dingle-Berry formula~\eqref{DBD} or its generalization~\eqref{conmat}. This means that, we can have an approximate analytic solution for mode functions as basis functions and background quantities are known functions. Accordingly, we are able to obtain analytic approximations of physical quantities. This is rather important since one needs to integrate over $k$ to obtain e.g. energy density as shown in~\eqref{energy}. It would be generally difficult to perform the momentum integration unless exact solutions are known. If we are able to obtain the approximate solution for the superadiabatic basis, using the approximate analytic solutions, we can perform the $k$-integration analytically or semi-analytically. Therefore, the superadiabatic basis would be useful to obtain the approximation for various physical quantities. Nevertheless, we need to be careful the following points. First, the optimal adiabatic order depends on $k$. Typically, the optimal order becomes higher for larger momentum modes. Therefore, strictly speaking, one need to change the optimal order mode by mode. This is impractical. Fortunately, large momentum modes are less created and one may just take the optimal order for the mode that mainly contribute to the physical quantity. Such an approximation would not spoil the estimation. The second caution is that taking higher order adiabatic expansion and performing $k$-integration would be a difficult task since the higher-order adiabatic expansion contain more terms than that of lower orders. One may perform momentum integration numerically, but analytic calculation would be not so easy. Therefore, if we need analytic expression, we may use the approximate solution of $(\alpha_k(t),\beta_k(t))$ given by~\eqref{conmat} while choosing the simplest choice $W_k^{(0)}=\omega_k$, $V_k=0$. For such choices, $Q_k\to {\rm i}\omega_k$, and the analytic expression would be simplified. We note that such approximation corresponds to neglecting higher-order terms in adiabatic expansion, which can be justified in sufficiently adiabatic time region.\footnote{We expect the approximation is not reliable in particular around the Stokes line crossing, where the adiabaticity is maximally violated.} If one needs better approximation, one can evaluate e.g. energy density by substituting approximate solutions of $(\alpha_k(t),\beta_k(t))$ into \eqref{rhonp} and taking the optimal adiabatic order. In such a case, one may take semi-analytic approach as we mentioned above. 

\section{Adiabatic particle numbers in cosmological models}\label{adpart}
We apply the superadiabatic basis discussed in Sec.~\ref{adrev} to several cosmological models, which will clarify how the difference of the basis exhibits the behavior of the ``particle number''.
\subsection{A toy model: a simple time-dependent mass}\label{toy}
Let us consider the following effective frequency
\begin{equation}
    \omega_k^2(t)=k^2+m^2+\lambda\mu^4t^2,\label{toyf}
\end{equation}
where $\mu$ is a mass parameter. This simple model captures the broad resonance regime of preheating and the moving D-brane model~\cite{Kofman:1997yn,Kofman:2004yc}. Indeed, if we take $h(\phi)=\phi^2$ in \eqref{action} and approximate the field dynamics as $\phi\sim \mu^2 t$ while ignoring expansion of the Universe, we find \eqref{toyf}.

\begin{figure}[htbp]
\centering
\includegraphics[width=.4\textwidth]{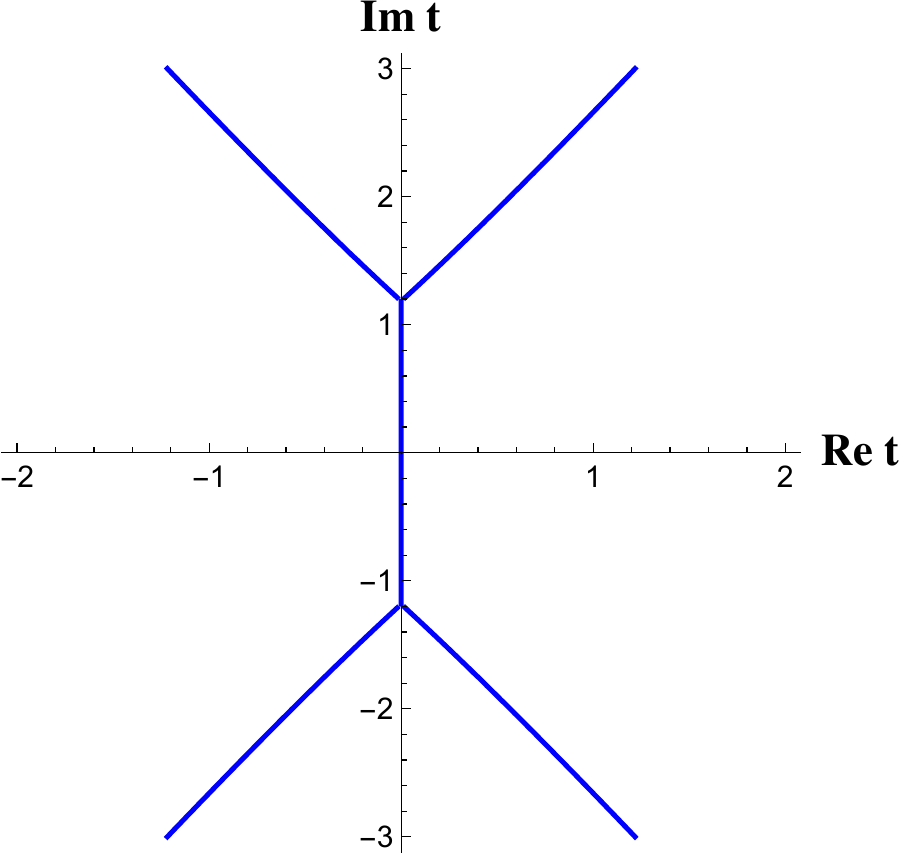}
\caption{Stokes lines on the complex $t$-plane in the model~\eqref{toyf}. We have taken $\lambda\mu^4=5,\  k^2+m^2=7$ in this figure. At each turning point, three Stokes lines emanate, and the one crossing real axis connects two turning points. }\label{fig;stokes}
\end{figure}

In order to estimate the amount of particle production, we need to discuss Stokes phenomenon in this setup. The turning points, at which $\omega_k^2=0$, can be found at
\begin{equation}
    t_c= {\rm i}\frac{\Omega_k}{\sqrt{\lambda}\mu^2},
\end{equation}
and its conjugate $\bar{t}_c$, where $\Omega_k=\sqrt{k^2+m^2}$. Since $\omega_k(t)$ is a real function of $t$ on the real axis, the Stokes line crossing the real axis at $t=0$ connects two turning points. Note that three Stokes lines emanate from each turning point, and two of them do not cross the real axis and go to infinity. (See Fig.~\ref{fig;stokes}.)

Using the Dingle-Berry formula~\eqref{BDf}, we can estimate the time dependence of the particle number when we take the optimal adiabatic order. Let us explicitly show the quantities necessary to obtain an analytic formula in the following. The amount of the particle production is determined by the phase integral along the Stokes line connecting two turning points~\eqref{fk0}, which is given by
\begin{equation}
    F_k^{(0)}={\rm i}\int_{t_c}^{\bar{t}_c}\omega_k dt=\frac{\pi \Omega_k^2}{2\sqrt{\lambda}\mu^2},
\end{equation}
where we have chosen the phase of $\omega_k$ so that $F_k^{(0)}>0$. The Stokes line crosses real axis at $t=s_c=0$. Then, we can approximate $F_k(t)$~\eqref{fkt} as
\begin{eqnarray}
    F_k(t)&=&2{\rm i}\left(\int_{t_c}^{s_c}+\int_{s_c}^t\right)\omega_k(t')dt'\nonumber\\
    &=& F_k^{(0)}+2{\rm i}\int_{s_c}^t\omega_k(t')dt'\nonumber\\
    &\sim&F_k^{(0)}+2{\rm i}\omega_k(s_c)(t-s_c)\nonumber\\
    &=&\frac{\pi \Omega_k^2}{2\sqrt{\lambda}\mu^2}+2{\rm i}\Omega_kt.
\end{eqnarray}
This expression is valid around $t\sim s_c(=0)$. With this expression, we find $\sigma_k$~\eqref{sigma} as
\begin{equation}
    \sigma_k(t)=\frac{{\rm Im}F_k(t)}{\sqrt{2{\rm Re}F_k(t)}}\sim \frac{2\lambda^{1/4}\mu t}{\sqrt{\pi}}.
\end{equation}
Thus, for the optimal truncation order, $\beta_k(t)$ would behave as
\begin{equation}
    \beta_k(t)\sim \frac{\rm i}{2}e^{-\frac{\pi \Omega_k^2}{2\sqrt\lambda \mu^2}}{\rm Erfc}\left(-\frac{2\lambda^{1/4}\mu t}{\sqrt\pi}\right),\label{massb}
\end{equation}
and also the time-dependent particle number is given by
\begin{equation}
    n_k\sim \frac{1}{4}e^{-\frac{\pi \Omega_k}{\sqrt\lambda \mu^2}}\left({\rm Erfc}\left(-\frac{2\lambda^{1/4}\mu t}{\sqrt\pi}\right)\right)^2.\label{toyDB}
\end{equation}
We find that this simple formula actually captures the behavior of the particle number evolution at the optimal order with the natural choice $V_k=-\frac{\dot{W}_k^{(j)}}{2W_k^{(j)}}$. We note that the form of $\beta_k$ in \eqref{massb} actually exhibits its non-perturbative property in the coupling constant $\lambda$. As we expected, $\beta_k$ decays exponentially as $k$ increases, and does not lead to UV divergences. 

\begin{figure}[htbp]
\centering
\includegraphics[width=.9\textwidth]{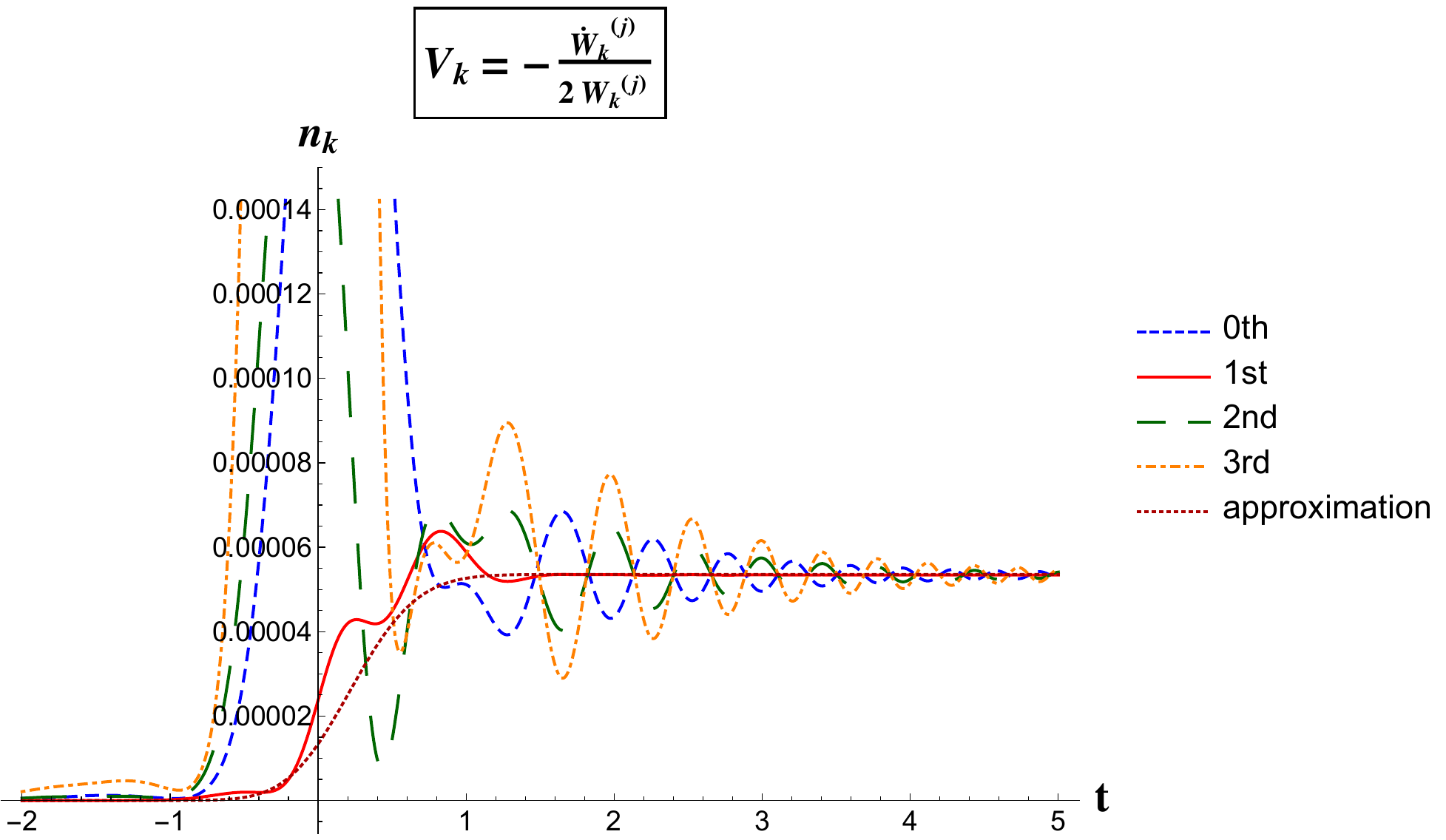}
\caption{The time dependence of the adiabatic particle number for different adiabatic orders $(j=0,1,2,3)$ in the model~\eqref{toyf} with the ``natural choice'' of $V_k$. We have taken $\lambda\mu^4=5,\  k^2+m^2=7$ in this figure. The red solid curve ($j=1$) shows a relatively smooth transition from zero to non-vanishing value. The dark red dotted line is \eqref{toyDB}, which shows good agreement with the first ($j=1$) adiabatic expansion (red solid line). Others ($j=0,2,3$) show large oscillatory behavior around Stokes line crossing $t=0$. Nevertheless, all converges to the same value for large $t$.}\label{fig;adv}
\end{figure}

\begin{figure}[htbp]
\begin{center}
\begin{tabular}{c}
\begin{minipage}{0.5\hsize}
\begin{center}
\includegraphics[width=.9\textwidth]{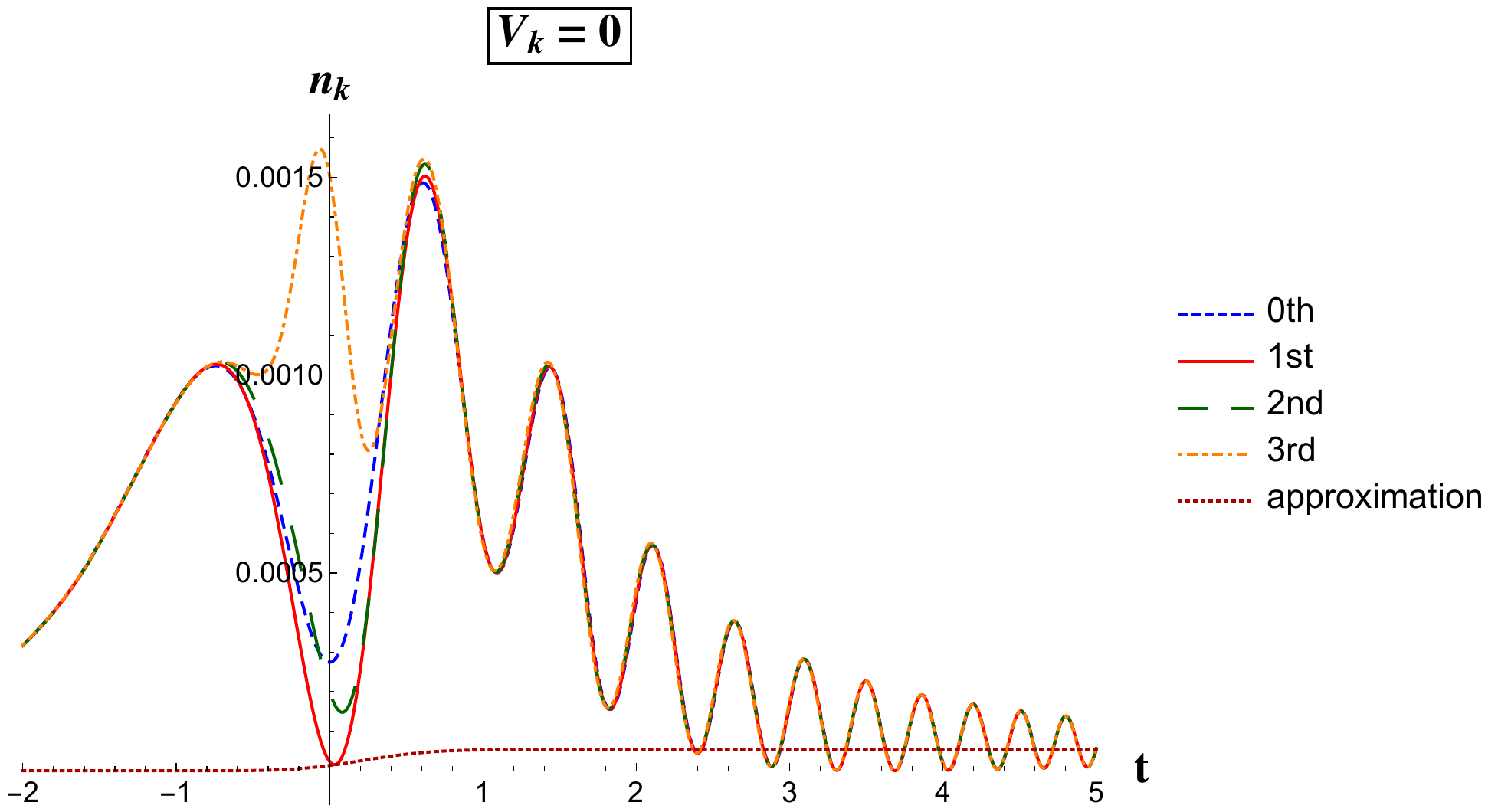}
\end{center}
\end{minipage}
\begin{minipage}{0.5\hsize}
\begin{center}
\includegraphics[width=.9\textwidth]{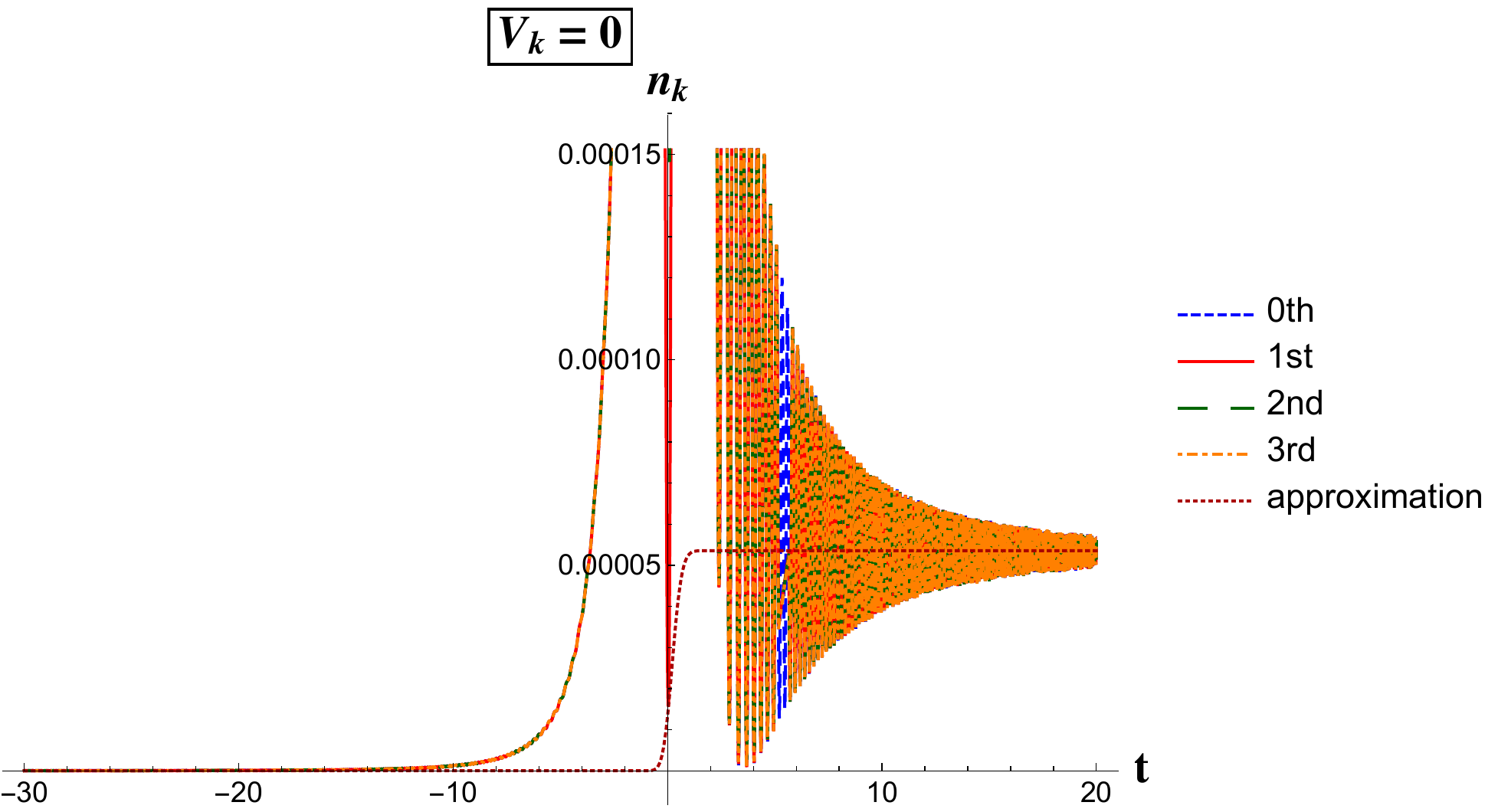}
\end{center}
\end{minipage}
\end{tabular}
\caption{The time dependence of adiabatic particle number for different adiabatic orders $(j=0,1,2,3)$ in the model~\eqref{toyf} with the standard (WKB) choice $V_k=0$. The left panel shows the behavior around $t=0$, and the right panel is the long time behavior. We have taken the same parameters as Fig.~\ref{fig;adv}. We find neither of adiabatic orders show smooth transition. Nevertheless, all eventually asymptotes to the same value consistent with Berry's formula. The dark red dotted line corresponds to \eqref{toyDB}.}
\label{fig;ad}
\end{center}
\end{figure}

For illustration, we show the numerically derived time-dependent particle number at different adiabatic orders with either $V_k=-\frac{\dot{W}_k^{(j)}}{2W_k^{(j)}}$ or $V_k=0$ in Fig.~\ref{fig;adv} (for natural choice of $V_k$), and Fig.~\ref{fig;ad} (for $V_k=0$).

As we clearly see from the figures, the first adiabatic expansion with the natural choice of $V_k$ shows a smooth transition, which is in good agreement with the approximate formula~\eqref{toyDB}. We also confirm that the asymptotic value of $n_k$ converges to the same, irrespective of the adiabatic order or the choice of $V_k$. Therefore, as long as we are interested in the final values of produced particle number, the different choices of either adiabatic orders $j$ or the choice of $V_k$ would not matter. Nevertheless, it would be appropriate to define a particle with $j=1$ and the natural choice of $V_k$ within this parameter set, which yields ``stable'' particle number. We note that, according to Berry's estimate of the optimal order~\eqref{jest}, our parameter choice is optimal at $j=1$, which is consistent with our result.\footnote{We should note that we find the numerical value $\frac12 (F^{(0)}_k-1)\sim 1.96$, which marginally indicates $j=1$ as the optimal order. This might be because the value of $F_k^{(0)}$ is not so large for our parameter choice, whereas Berry's derivation of the optimal order assumes a large value of $F_k^{(0)}$. We may understand the small discrepancy between \eqref{toyDB} and the first adiabatic order result in Fig.~\ref{fig;adv} in the same way.}

Thus, we have confirmed that with an appropriate choice of the adiabatic order and $V_k$, we can have a proper adiabatic particle, which shows (1) smooth particle production around Stokes line crossing, (2) stable (non-oscillatory) behavior of particle number. Besides that, the proper adiabatic particle number or $\beta_k(t)$ is well approximated by a very simple formula~\eqref{toyDB}. It is worth emphasizing that if we can have an analytic approximation for $\beta_k$, we are able to calculate physical quantities such as energy density (see~\eqref{energy}). Note that $\alpha_k$ should satisfy the normalization condition $|\alpha_k|^2-|\beta_k|^2=1$, and once $|\beta_k|$ is fixed, the absolute value of $\alpha_k$ can be determined. Therefore, we would be able to have approximate formulas for various physical quantities with a simple formula~\eqref{toyDB}.

In the following, we will reconsider the cases where multiple particle production events are relevant and interference of events affect the time-dependent particle number.

\subsection{Narrow and broad resonance regime in the oscillating scalar background}\label{scalarosc}
We discuss the production of a scalar particle $\chi$ coupled to an oscillating scalar field $\phi$ in Minkowski background, which is a toy model of preheating after inflation. More explicitly we consider $h(\phi)=\phi(t)=A\cos (Mt)$, where $A$ is a dimensionful constant. Then, the effective frequency of $\chi$ for a mode with momentum $k$ is given by
\begin{equation}
    \omega_k^2=k^2+m^2+\lambda A \cos (Mt).\label{omgres}
\end{equation}
Depending on the set of parameters, we have three different regime: For $\lambda A>k^2+m^2$, there are time intervals when the effective frequency becomes pure imaginary, namely a tachyonic instability regime. We will not discuss this regime here.\footnote{In~\cite{Hashiba:2021npn}, Stokes phenomenon in the tachyonic regime is discussed with in $R^2$ inflation model. In this case, the Stokes line appears along the real time axis. See also~\cite{Dufaux:2006ee}.} For $\lambda A\ll k^2+m^2<M^2$, the modes in a certain momentum range experiences the so-called narrow resonance regime, which will be explained below. For $M^2\ll \lambda A<k^2+m^2$, we find the broad resonance regime. The broad resonance, in particular, cannot be understood from a perturbative scattering viewpoint, since the decay of $\phi$ to $\chi$ is kinematically forbidden.

Let us discuss the narrow resonance case $\lambda A\ll k^2+m^2<M^2$. For most momentum modes, the amount of particle production is quite small. Nevertheless, the modes in the resonance band around $4(k^2+m^2)\sim M^2$ shows particle number exponentially growing in time. One of the reasons to reconsider the narrow resonance regime here is for the following question: Can we describe the narrow resonance in the same way as the broad resonance? It is well known that resonance in this model can be understood in terms of Mathieu equation: The mode equation of $v_k$ is rewritten as
\begin{equation}
    v''_k+(A_k-2q \cos(2z))v_k=0,
\end{equation}
where $A_k=4(k^2+m^2)/M^2$, $q=2\lambda A/M^2$, and $z=Mt/2 +\pi$. The first resonance band condition is $1-q<A_k <1+q$~\cite{Kofman:1997yn}. The center of the resonance band $k^2+m^2\sim M^2/4$ seems similar to the possible kinematic configuration of $\phi\to \chi\chi$ decay channel. It is known that the mode function inside the resonance band grows as~\cite{Kofman:1997yn}
\begin{equation}
    v_k(t)\propto \exp\left(\frac12 qz\right)\propto \exp\left(\frac{\lambda At}{2M}\right),
\end{equation}
and therefore the particle number would grow as $n_k\propto \exp (\lambda At/M)$. Therefore, we interpret $\Gamma=\lambda A/M$ as the production rate of $\chi$. The narrow resonance can be understood from the perturbative decay of inflaton to $\chi$ with Bose enhancement effects (see e.g.~\cite{Mukhanov:2005sc}).\footnote{We note however that the production rate is quite different from that given by a perturbative analysis. For clarification of this point, we give a perturbative analysis of $\chi$-production in appendix~\ref{pertWKB}.}

On the other hand, in the broad resonance regime, produced particle number is expressed in terms of non-analytic functions of a coupling constant $\lambda$ as it can be understood from the simplified model in Sec.~\ref{toy}. If we try to describe both the narrow and broad resonance regime, it would mean that we need to treat perturbative and non-perturbative contributions simultaneously. We will show that this is possible, namely, we are able to describe both regime in a unified way. This is not so surprising since the both regime can be described by a single differential equation, Mathieu equation. Nevertheless, it is important to confirm that Stokes phenomenon captures both perturbative and non-perturbative effects. 

\begin{figure}[htbp]
\centering
\includegraphics[width=.4\textwidth]{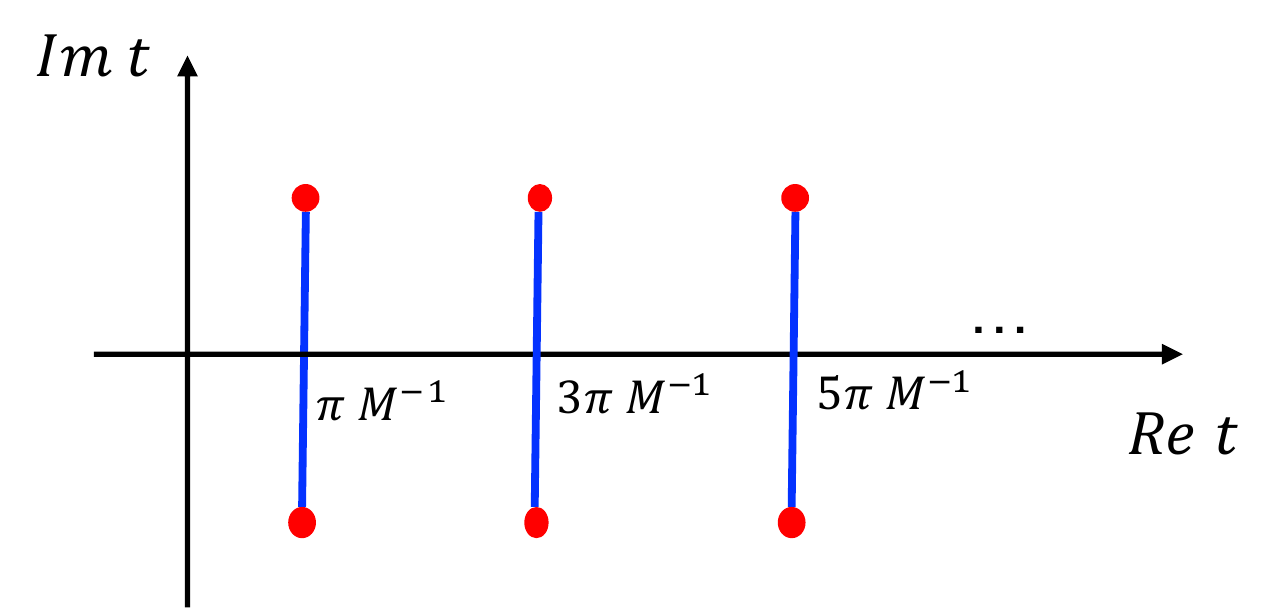}
\caption{Schematic picture of Stokes lines (blue lines) and turning points (red dots) of the effective frequency~\eqref{omgres}. Note that we draw only the relevant Stokes lines crossing the real axis, and there are three Stokes lines emanating from each turning point. }\label{fig;presto}
\end{figure}

Let us discuss the Stokes phenomenon in this model. Turning points at which $\omega_k^2(t_c)=0$ are 
\begin{equation}
    t_c=\frac1M \left((2n-1)\pi\pm{\rm i}\log (c+\sqrt{c^2+1})\right),
\end{equation}
where $c=(k^2+m^2)/(\lambda A)$, and $n$ is an arbitrary integer. The appearance of $n$ represents the repeated particle production events, and we will take $n=1$ to be the first event. A turning point and its complex conjugate are connected by a Stokes line that crosses the real $t$ axis. We illustrate the Stokes line structure in this model in Fig.~\ref{fig;presto}. The amount of the particle produced at each Stokes line crossing is determined by
\begin{equation}
    F^{(0)}_k={\rm i}\int_{\bar{t}_c}^{t_c}dt\omega_k=-4{\rm i}\frac{\sqrt{\lambda A}}{M}\sqrt{c-1}E\left(\frac{\rm i}{2}\log(c+\sqrt{c^2+1})\bigl| -\frac{2}{c-1}\right),
\end{equation}
where $E(a|x)=\int_0^a\sqrt{1-x^2\sin^2\theta}d\theta$ is the incomplete elliptic integral of the second kind. The interference effect among each Stokes line crossing is characterized by the phase shift 
\begin{equation}
    \theta=\int_0^{2\pi}\omega_kdt=\frac{2\sqrt{k^2+m^2}}{M\sqrt{c}}\left(\sqrt{c-1}E\left(-\frac{2}{c-1}\right)+\sqrt{c+1}E\left(\frac{2}{1+c}\right)\right),
\end{equation}
where $E(x)=E(\frac\pi2|x)$ is the complete elliptic integral of the second kind. Note that for the large $c$ limit, we find $\theta\to\frac{2\pi\sqrt{k^2+m^2}}{M}$, and for $c\to1$, $\theta\to 0$. Therefore, at the first resonance band $k^2+m^2\sim M^2/4$, the phase factor becomes $\theta\sim \pi$. Our approximate connection matrix~\eqref{conmat} at $n$-th Stokes line crossing becomes
\begin{equation}
    C_n=\left(\begin{array}{cc}\sqrt{1+|E_{k,n}(t)|^2}& {\rm i}E_{k,n}(t)\\ -{\rm i}(E_{k,n}(t))^*& \sqrt{1+|E_{k,n}(t)|^2}\end{array}\right),\label{resapp}
\end{equation}
where
\begin{equation}
    E_{k,n}(t)=\frac12 e^{2{\rm i}n\theta}e^{-F^{(0)}_k}{\rm Erfc}(-\sigma_{k,n}(t))
\end{equation}
and
\begin{equation}
    \sigma_{k,n}(t)\sim \frac{2\sqrt{k^2+m^2-\lambda A}}{\sqrt{2F_k^{(0)}}}(t-(2n-1)\pi).
\end{equation}
Note that due to the periodicity of the background, the phase factor is simply given by the multiple of $e^{{\rm i}\theta}$. One can obtain $(\alpha_k(t),\beta_k(t))$ including particle production up to the $n$-th event as
\begin{equation}
    \left(\begin{array}{c}\alpha_k(t)\\ \beta_k(t)\end{array}\right)=\prod_{j=1}^n C_j\left(\begin{array}{c}1\\ 0\end{array}\right)\label{oscapp},
\end{equation}
where we have assumed the adiabatic vacuum condition $(\alpha_k(0),\beta_k(0))=(1,0)$. One may use the simpler approximation formula~\eqref{DBD} if $n_k<1$, but it fails for the case $n_k>1$. The formula~\eqref{oscapp} is all we need for evaluation of the particle production. One can immediately obtain the particle number $n_k=|\beta_k|^2$ from the formula~\eqref{oscapp}.

We give some general remarks about our approximate formula~\eqref{resapp}. The formula shows time-dependence in good agreement with numerically solved mode equation in various parameter regions. However, it fails to reproduce the numerical result exactly. There are several reasons for the discrepancy. Our conjectured formula~\eqref{conmat} is based on the combination of Dingle-Berry's error function formula and the connection formula of the parabolic cylinder function. This approximation would be applicable if each Stokes line crossing the real axis is well separated from others. If we try to improve the connection matrix formula, we need to perform more careful analysis of Stokes phenomenon using e.g. symmetry relations~\cite{Froman1,Kutlin:2017dpe}.\footnote{It would also be important to modify Berry's derivation of the error function formula in order to find the time dependence of the Bogoliubov coefficients $\alpha_k(t),\beta_k(t)$. To our best knowledge, the generalization of Berry's formula in the presence of multiple turning points is not known.} Nevertheless, our simple formula~\eqref{conmat} would be useful for practical purposes and reproduce correct behavior in most cases. 
\begin{figure}[htbp]
\centering
\includegraphics[width=.7\textwidth]{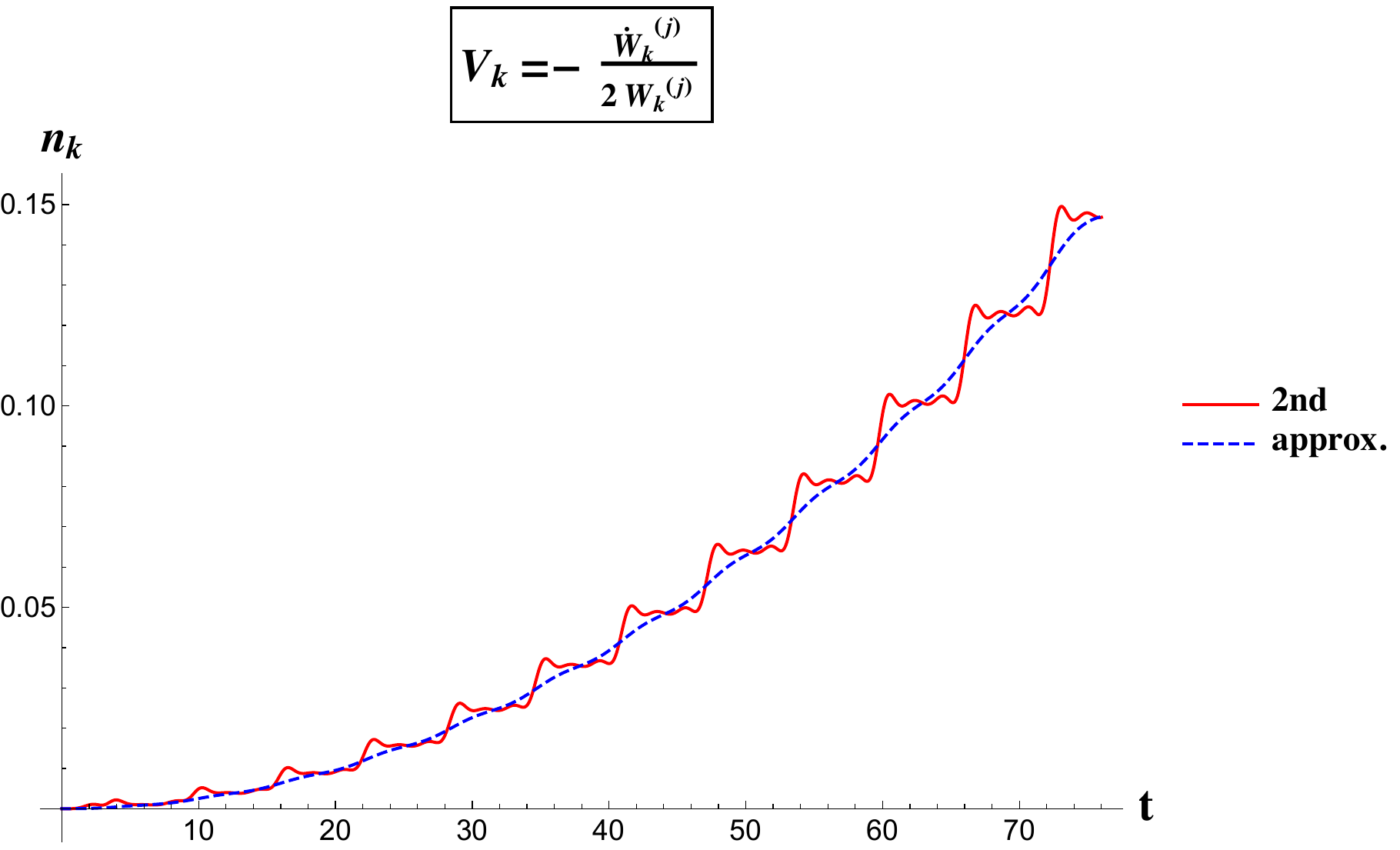}
\caption{Time-dependent particle number for the narrow resonance case. We have taken $k^2+m^2=1/3.99, M=1, \lambda A= 0.01$, and the adiabatic order is taken to be the second order with the natural choice of $V_k$ for the numerical result (red solid curve). Time $t$ is in unit of $M^{-1}$. We should note that we put the overall factor $1/1.4$ to the approximate formula, namely, the blue dashed curve is $|\beta_k(t)|^2/1.4$ with $\beta_k(t)$ in \eqref{oscapp}.}\label{fig;narrow}
\end{figure}
\begin{figure}[htbp]
\centering
\includegraphics[width=.7\textwidth]{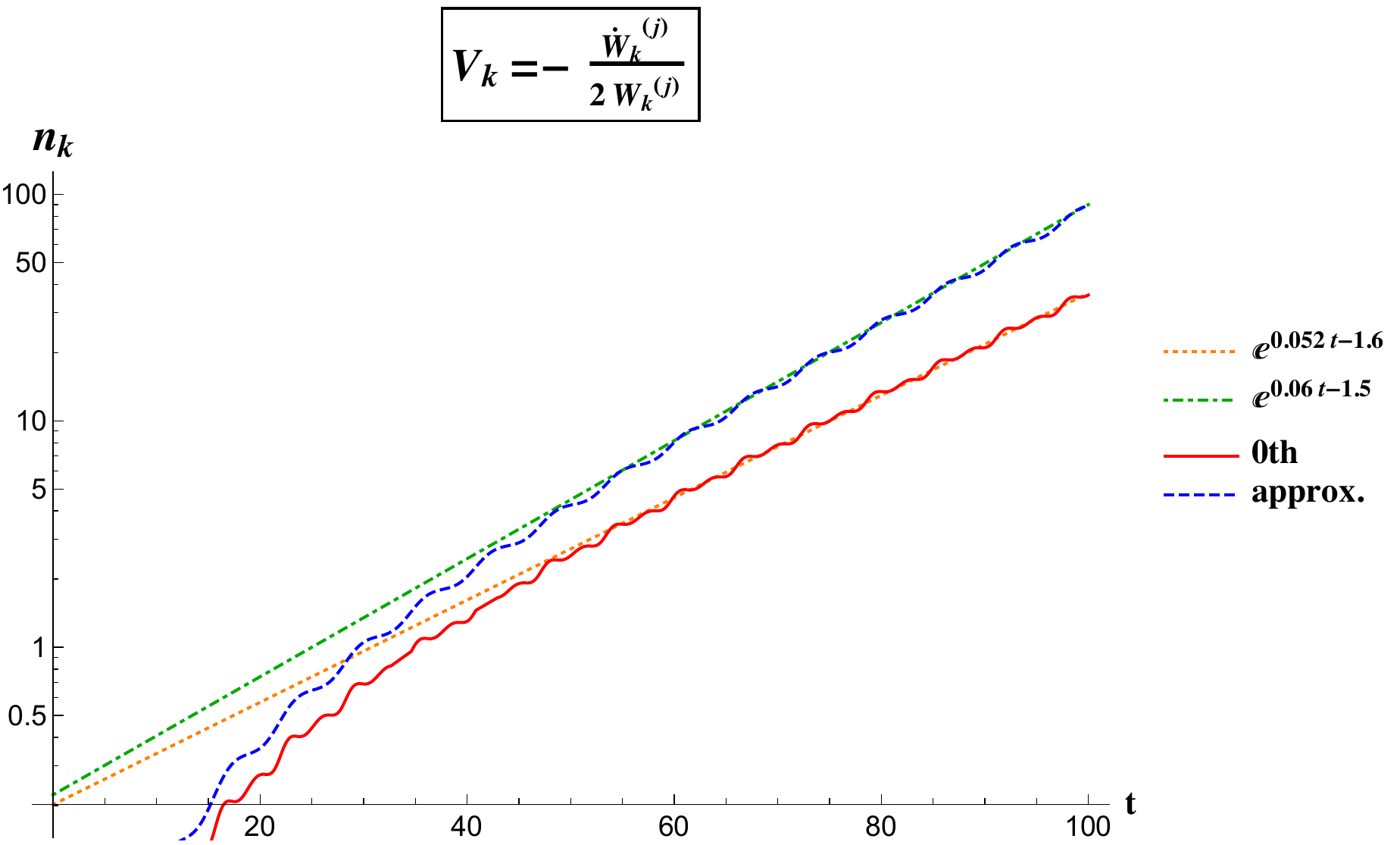}
\caption{Long time behavior of narrow resonance regime. In this figure, we take a parameter set different from that of Fig.~\ref{fig;narrow}: $\lambda A=0.05,\ k^2+m^2=1/3.99,\ M=1 $. The time $t$ is in unit of $M^{-1}$. To see the exponential growth behavior $n_k\propto e^{\lambda A t/M}$, we also draw the exponential functions for comparison. Our approximate formula shows slightly larger production rate than the numerical result. Nevertheless, we find the expected growth rate behavior $\Gamma\sim \lambda A/M$.}\label{fig;longnarrow}
\end{figure}
\begin{figure}[bhtp]
\centering
\includegraphics[width=.7\textwidth]{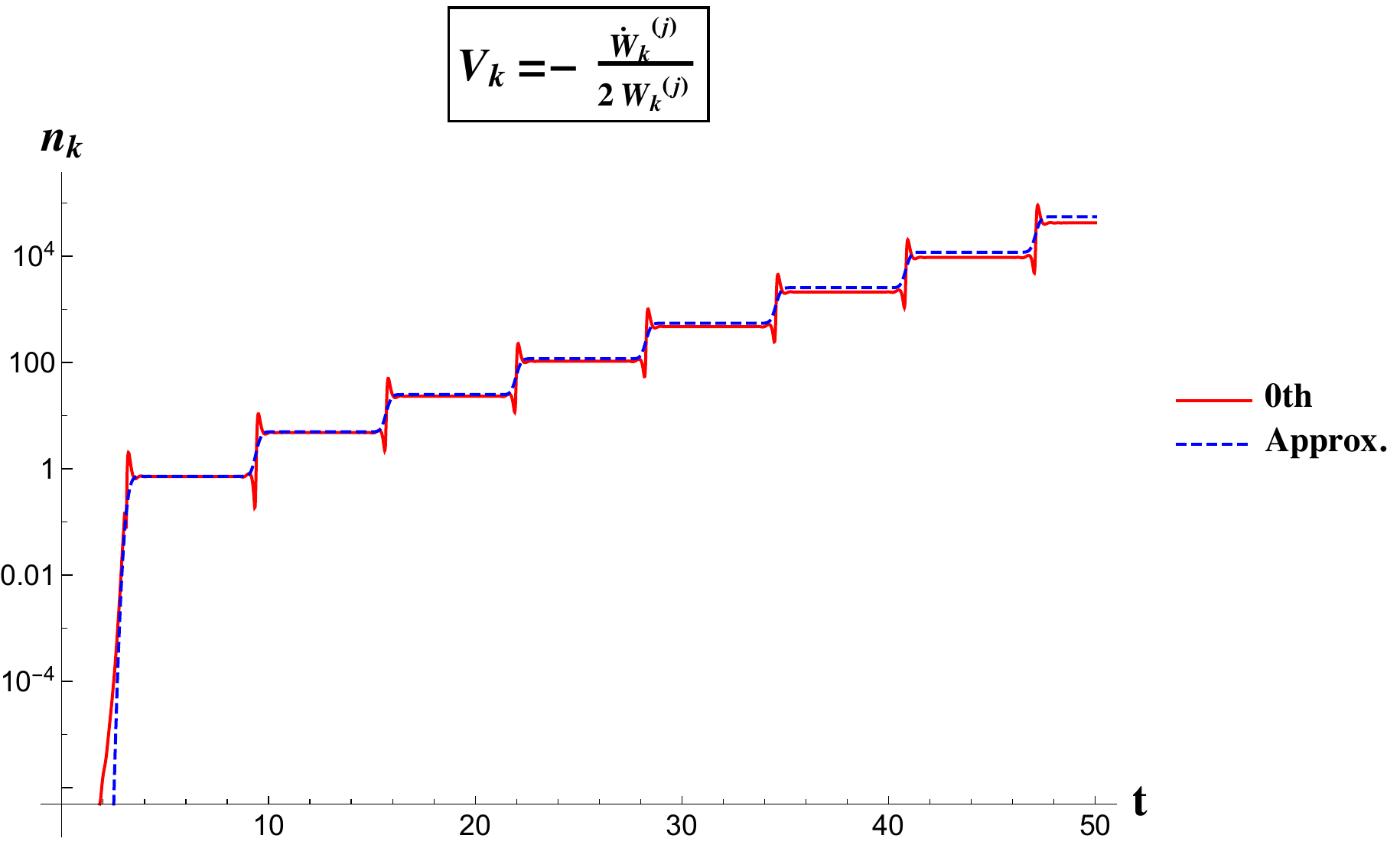}
\caption{Time-dependent particle number for the broad resonance case. We have taken $k^2+m^2=191, M=1, \lambda A= 190$, and for numerical result (red solid curve) the adiabatic order is the zero-th order with the natural choice of $V_k$. The blue dashed line is that evaluated by~\eqref{oscapp}.}\label{fig;broad}
\end{figure}
\begin{figure}[htbp]
\centering
\includegraphics[width=.7\textwidth]{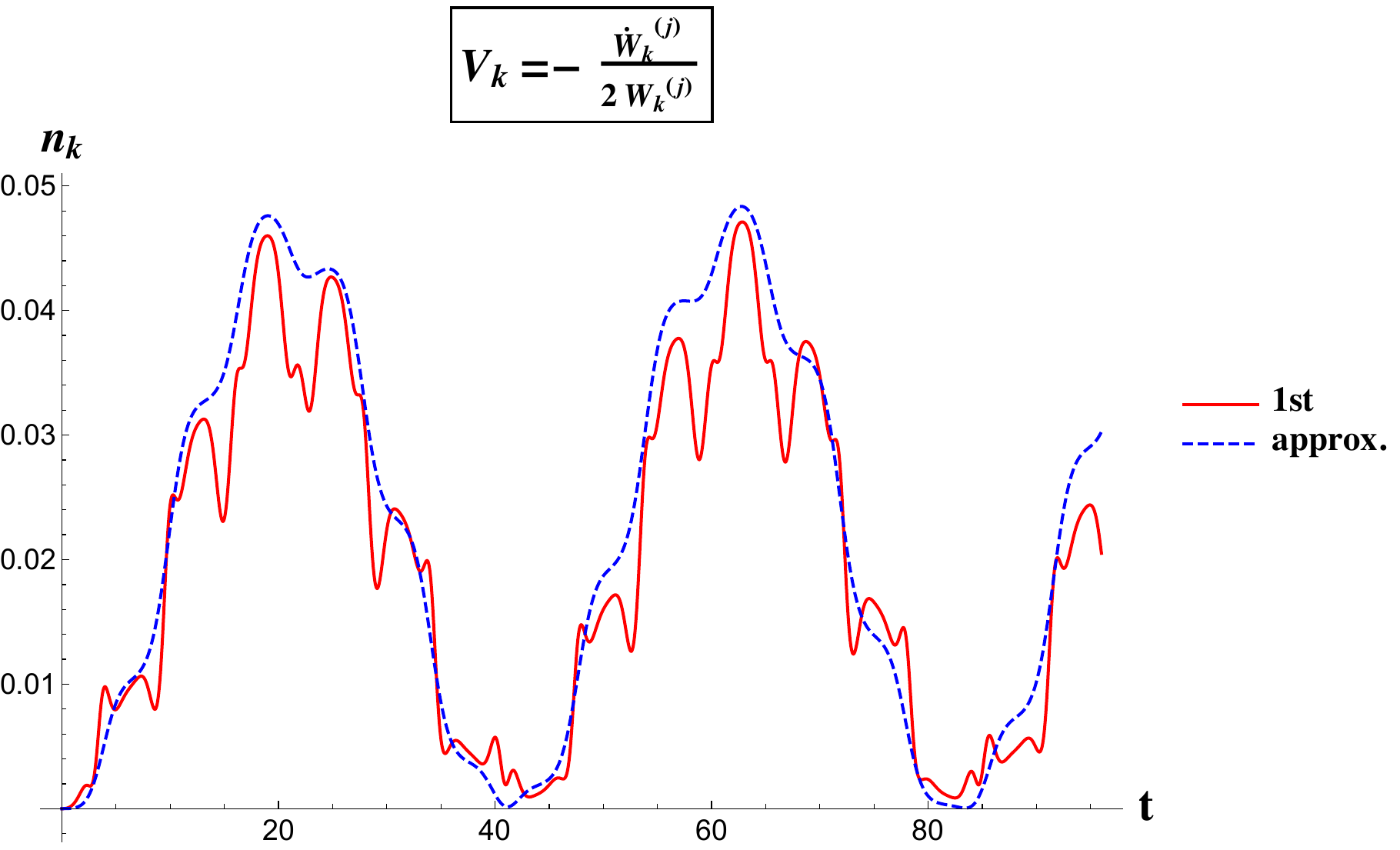}
\caption{Time-dependent particle number for a non-resonant case. We have taken $k^2+m^2=1/2.99, M=1, \lambda A= 0.05$, and the adiabatic order is the first order with the natural choice of $V_k$. The red solid curve is numerical result and the blue dashed line is that evaluated by~\eqref{oscapp}.}\label{fig;int1}
\end{figure}

Let us show the comparison between $n_k=|\beta_k(t)|^2$ with our approximate formula~\eqref{oscapp} and that of numerical solutions. In evaluating particle number numerically, we used Emarkov-Milne equation reviewed in appendix~\ref{EMeq}. In Fig.~\ref{fig;narrow}, we show the narrow resonance case. We see plateau regions for the numerical solution, but such regions do not appear in our approximate formula. This is because the width of the error function is too large and separated particle production events are smoothly connected. Nevertheless, we see the growth behavior consistent with the numerical solution, up to an $\mathcal{O}(1)$ overall factor. We show a simple modification in appendix~\ref{narrowmod}, which shows a better fitting. We also show the long time behavior of the narrow resonance regime in Fig.~\ref{fig;longnarrow} (with a slightly different parameter set). We find the exponential growth rate expected from the instability band analysis is approximately reproduced. We would like to clarify the origin of the growth. As we mentioned earlier, the phase factor $\theta\sim \pi$ in this regime, and therefore, the phase factor appearing in the connection matrix does not give any interference effects. Effectively, after $n$ events, we just multiply the same matrix for $n$ times. One can easily confirm that such multiplication of matrices leads to exponential growth of $|\beta_k|$. This is the reason for the narrow resonance from our viewpoint. Such a viewpoint seems different from the instability analysis of Mathieu equation or Floque theory~\cite{Amin:2014eta}.

In Fig.~\ref{fig;broad}, the broad resonance case is shown. In this case, we find a very good agreement between the numerical solution and our approximate formula. We also show a non-resonant case in Fig.~\ref{fig;int1}. We find that our approximate formula looks an envelope of the numerical result. Thus, our simple formula is in good agreement with numerical evaluation of particle number in different parameter regions. However, we should emphasize a possible problem in our approximation. In the case of resonant particle production, the particle number grows exponentially, and therefore, small deviation at earlier time could be enhanced. Indeed, as shown in Fig.~\ref{fig;longnarrow}, the growth rate of our analytic approximation shows a slightly larger value than that of numerical result or expectation from the instability band analysis. This discrepancy results in large deviation due to the exponential dependence. As we quoted in appendix~\ref{univform}, it is not so easy to improve this point. Therefore, one needs to be careful about such an issue when applying our analytic formulas to models of explosive particle production. A similar issue appears in the preheating in expanding universe. As we will see in the next section, it is quite difficult to determine the interference effects among number of events. Therefore, we are able to find at most the estimate with a stochastic interference effects, which is known from earlier work~\cite{Kofman:1997yn}. Independently of these issues, one has to take into account backreaction to the background field, which cannot be neglected if the particle number is exponentially large. Then, one needs to perform e.g. lattice simulations~\cite{Khlebnikov:1996mc,Khlebnikov:1996wr,Prokopec:1996rr,Khlebnikov:1996zt} to take the backreaction into account. Even if one tries to use Stokes phenomenon analysis, the backreaction needs to be taken into account since it changes the background dynamics and accordingly the structure of Stokes lines. Then, the interference effects would become stochastic one. Therefore, a simple exponential growth would not be realized in reality. 

It is also important to check the UV behavior of $\beta_k(t)$. As we showed, the absolute value of $\beta_k$ is characterized by $e^{-F_k^{(0)}}$. Let us see how $F_k^{(0)}$ behaves for high momentum region, which corresponds to $c\to\infty$. From the definition of the incomplete Elliptic integral of the second kind, we may approximate $F_k^{(0)}$ as
\begin{equation}
    F_k^{(0)}\sim-4{\rm i}\frac{\sqrt{\lambda A}}{M}\sqrt{c}\int_0^{\frac{\rm i}{2}\log(2c)}d\varphi=\frac{2\sqrt{\lambda A}}{M}\sqrt{c}\log(2c).
\end{equation}
Noting $c=(k^2+m^2)/(\lambda A)$, we find exponential decay of $|\beta_k|$ for large $k$. Thus, the UV divergence does not appear e.g. in $\langle\rho_{\rm np}\rangle$ in \eqref{rhonp}. The UV finiteness is reasonable since the high momentum modes would not feel oscillation of mass effectively.

We have shown that the narrow and broad resonance regime are described in a unified way. In the broad resonance regime, the particle production rate at one Stokes line crossing is approximated by the model in~Sec.~\ref{toy}, which is non-perturbative in the coupling constant $\lambda$.\footnote{Around the minimum of the potential $Mt_0=\pi/2+2n\pi$, one can expand $\cos Mt\sim-1+ M^2(t-t_0)^2/2$ and find the correspondence to the model in Sec.~\ref{toy}. Such an approximation is used in the analysis of broad resonance in preheating~\cite{Kofman:1997yn}.} On the other hand, narrow resonance regime shows a perturbative behavior with respect to $\lambda$, since the growth rate is $\Gamma\propto \lambda$. The interpolation between perturbative and non-perturbative regimes is observed in the context of Schwinger effect~\cite{Dunne:2005sx,Basar:2015xna}. The weak field limit or multi-photon regime would correspond to the narrow resonance and the Schwinger limit/strong field limit corresponds to the broad resonance. Clarifying the correspondence would be interesting from theoretical viewpoint, and we will leave it for future study.

\subsection{Preheating in expanding universe}\label{preheating}
In this section, we discuss preheating in the expanding Universe. We take in \eqref{action} $a(t)=(1+H_{\rm ini}t)^{2/3}$, $h(\phi)=\phi^2$, $\phi(t)=A\sin(Mt)/t$, where $H_{\rm ini}$ denotes the initial value of the Hubble parameter at the beginning of preheating stage, and $A$ is a dimensionless constant.\footnote{The coupling constant $\lambda$ in this case is dimensionless.} For simplicity, we take $H_{\rm ini}=M$, and note that this setup does not describe the beginning of inflaton oscillation precisely, but is a good approximation in later time. As we will see, the most important dynamics is the oscillating part and the details of the rest would not affect much. We here neglect the terms including Hubble parameter and its time derivative in \eqref{genomg}. The effective frequency of $\chi$ with momentum $k$~\eqref{genomg} is given by
\begin{equation}
    \omega_k^2=\frac{k^2}{(1+Mt)^{4/3}}+m^2+\frac{\lambda A^2}{t^2}\sin^2 (Mt).\label{omgpre}
\end{equation}
Due to the expansion of the Universe, the amplitude of inflaton oscillation decays in time, and therefore, particle production is dominated by the early stage. We assume $A\gg1$, and then the oscillating term dominates the time-dependence of~\eqref{omgpre}. 

In order to make an estimate of particle production rate, let us discuss the Stokes phenomenon in this model. When $A\gg1$, the following approximation used in~\cite{Hashiba:2021npn} can be applied: Since the dominant term in \eqref{omgpre} is the sinusoidal term, the turning point would be near $Mt\sim n\pi$, where $n$ is a positive integer. We assume the turning point to be $Mt_n=n\pi+\delta_n$. Unless $n\gg1$, we expect $|\delta_n|\ll1$, and therefore,
\begin{equation}
  \frac{k^2}{(1+n\pi)^{4/3}}+m^2+\frac{\lambda A^2}{(n\pi)^2}\sin^2 (\delta_n)\sim0, 
\end{equation}
or equivalently,
\begin{equation}
    \sin^2\delta_n\sim -c_n^2,
\end{equation}
where
\begin{equation}
   c_n^2\equiv \frac{(n\pi)^2}{\lambda A^2}\left(\frac{k^2}{(1+n\pi)^{4/3}}+m^2\right).
\end{equation}
\begin{figure}[htbp]
\centering
\includegraphics[width=.4\textwidth]{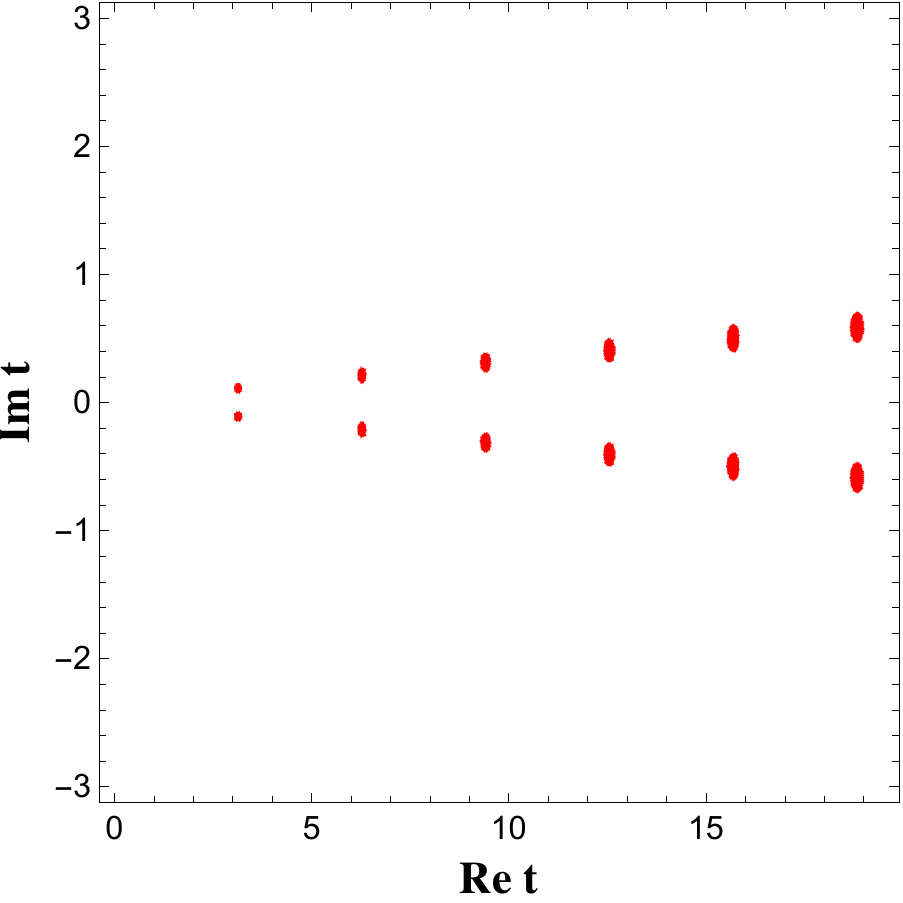}
\caption{Turning points of \eqref{omgpre} for $k=1,m=1,\lambda=1,A=900,M=1$. The red circles denote turning points. }\label{fig;tp}
\end{figure}
Therefore, $\delta_n\sim\pm{\rm i\ arcsinh}\ c_n$. As seen from this relation, the absolute value of the imaginary part of the turning points becomes larger as $n$ increases as illustrated in Fig.~\ref{fig;tp}. This behavior can be understood from the decay of the amplitude of the oscillator. The complex conjugate pairs of turning points are connected by Stokes lines crossing the real axis. We integrate the effective frequency along the Stokes line, and find
\begin{equation}
    F_{k,n}^{(0)}\sim \frac{{\rm i}\sqrt{\lambda}A}{n\pi }\int_{\bar{t}_n}^{t_n}\sqrt{c_n^2+\sin^2 Mt}dt=-\frac{2\sqrt{\lambda}A\rm i}{n\pi M}c_nE\left({\rm i}\ {\rm arcsinh}\ c_n \Bigl|-\frac{1}{c_n^2}\right).\label{Fpre}
\end{equation}
For $c_n\ll1$, we find 
\begin{equation}
    F_{k,n}^{(0)}\sim \frac{\sqrt{\lambda}Ac_n^2}{2nM},
\end{equation}
which can also be found from linear approximation, $\sin Mt\sim Mt$ in performing integration. Let us evaluate the amount of particle production at each Stokes line crossing. The $n$-th Stokes line crossing would give $\chi$-particle number density
\begin{equation}
 n_{k,n}\sim e^{-2F_{k,n}^{(0)}}~\sim \exp\left(-\frac{n\pi^2}{\sqrt{\lambda}AM}\left(\frac{k^2}{(n\pi+1)^{4/3}}+m^2\right)\right),\label{preapp}
\end{equation}
where we have assumed $c_n\ll1$.\footnote{One can check that this approximation works well until $c_n\sim 1$.} Note that this is not the particle number after $n$-th crossing, since we have to multiply the connection matrix at each Stokes line crossing as~\eqref{conmat}.

We comment on the choice of the adiabatic order in preheating setup. Due to decreasing amplitude, there are several particle production events with different values of $F_{k,n}^{(0)}$, which determines the optimal order according to Berry's estimate~\eqref{jest}. The leading contribution is the first Stokes line crossing $n=1$, and therefore, determining the optimal order from $F_{k,1}^{(0)}$ would be the best choice since later events only give small changes of particle number.\footnote{If the later contributions are not negligible, it would mean that their $F_{k,n}^{(0)}$ would be the same order as $F_{k,1}^{(0)}$, then the optimal order for them would be the same.} As naively expected, larger amplitude $A$ reduces $F_{k,1}^{(0)}$, and for explosive particle production case, the zero-th order is the optimal choice for the adiabatic basis. For heavy particle production cases discussed later, the higher adiabatic order is the optimal basis.

For the production of light particles, numbers of particle production events interfere with each other. More specifically, the phase factor $\theta_{k,n}$ in \eqref{Ek} describes the interference effects. However, the evaluation of $\theta_{k,n}$ requires very good accuracy, since $\mathcal{O}(1)$ error leads to completely different interference patterns. We could not find a good analytic formula with $O(1)$ accuracy. Besides that, when $F_{k,n}^{(0)}$ is very small until $n$ becomes very large, which is the case for explosive particle production, our approximate connection matrix formula~\eqref{conmat} is not a good approximation as we mentioned earlier. Therefore, it would be more practical to treat $\theta_{k,n}$ as stochastic parameters as the treatment in~\cite{Kofman:1997yn}. In this case, it is difficult to predict the time-dependent particle number accurately. Nevertheless, we show in appendix~\ref{stoch} that our connection formula~\eqref{conmat} can capture the behavior of explosive particle production qualitatively.

\begin{figure}[htbp]
\begin{center}
\begin{tabular}{c}
\begin{minipage}{0.5\hsize}
\begin{center}
\includegraphics[width=0.99\textwidth]{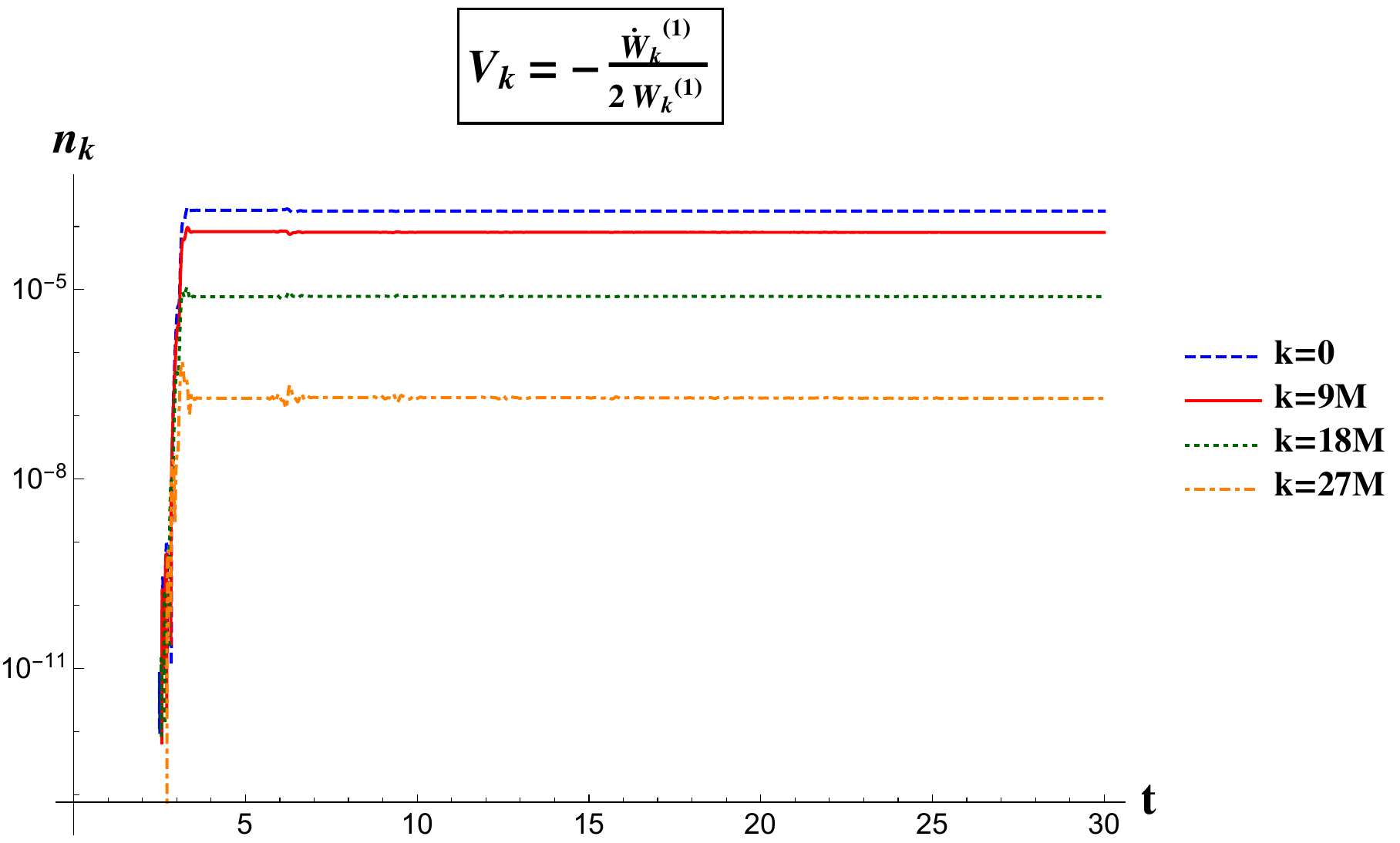}
\end{center}
\end{minipage}
\begin{minipage}{0.5\hsize}
\begin{center}
\includegraphics[width=0.99\textwidth]{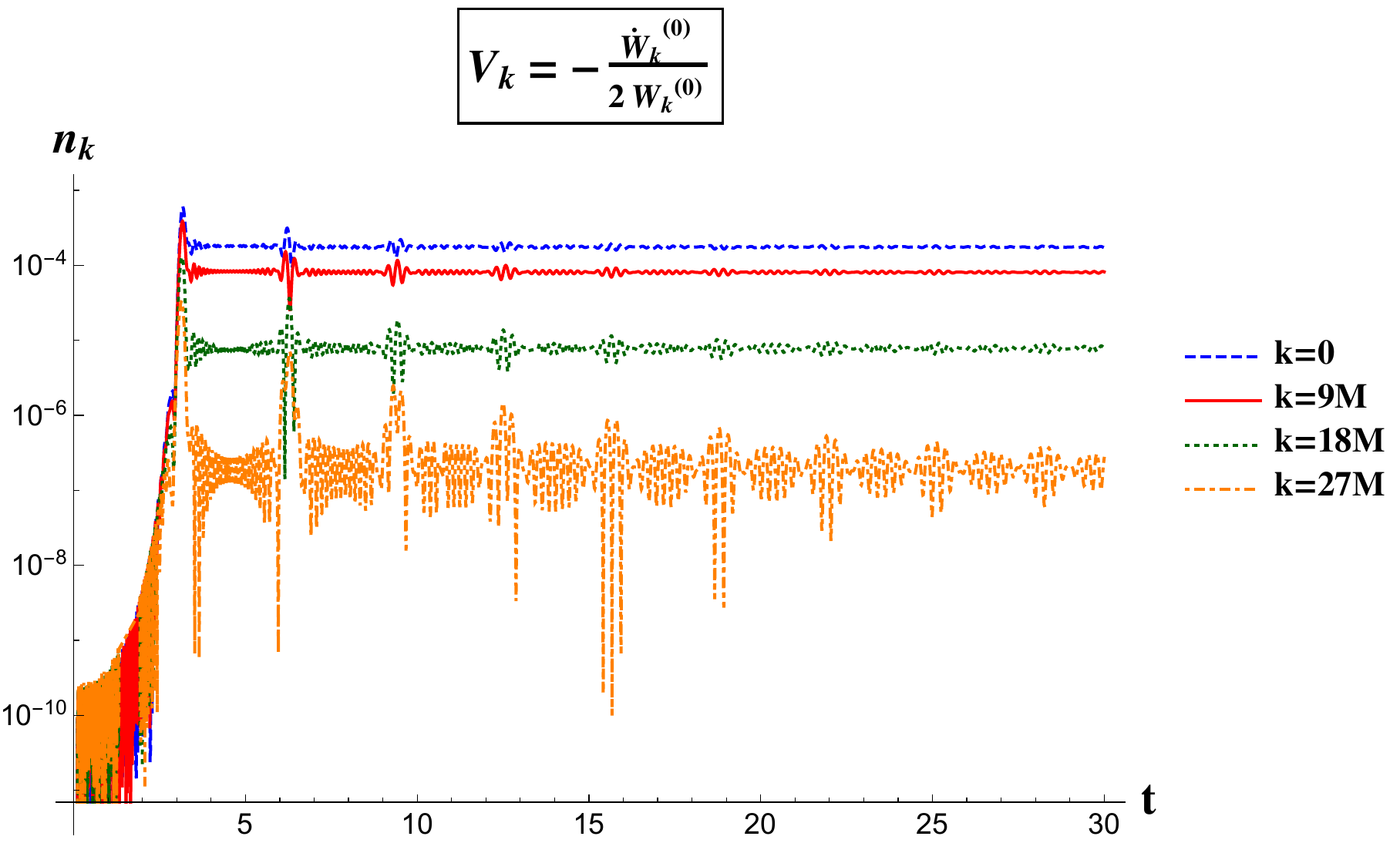}
\end{center}
\end{minipage}
\end{tabular}
\caption{Heavy particle production in expanding universe with the parameter set~\eqref{para}. Left panel shows the numerical solution with the first adiabatic order and the right panel is with the zero-th order. We find that the first adiabatic expansion gives stable particle number. We also find that only the first Stokes line crossing around $ t \sim \pi$ is responsible for the production. Note that the time $t$ is in unit of $M^{-1}$.}
\label{fig;massive}
\end{center}
\end{figure}

Let us consider a relatively simple case, where only the first event is responsible for particle production. For concreteness, we take the following parameters,
\begin{equation}
   \lambda=1,\ A=150,\ m=11.5M, \label{para}
\end{equation}
and we will take $M=1$, namely all the dimensionful quantities are in unit of $M$. Since $\chi$-particle is quite heavy, only the first Stokes line crossing is relevant for particle production. In Fig.~\ref{fig;massive}, we show the time-dependent particle number for different momentum modes, and also the difference between the zero-th and first adiabatic order. As clearly seen, the first order expansion give non-oscillatory ``particle number''. 
\begin{figure}[bhtp]
\centering
\includegraphics[width=.7\textwidth]{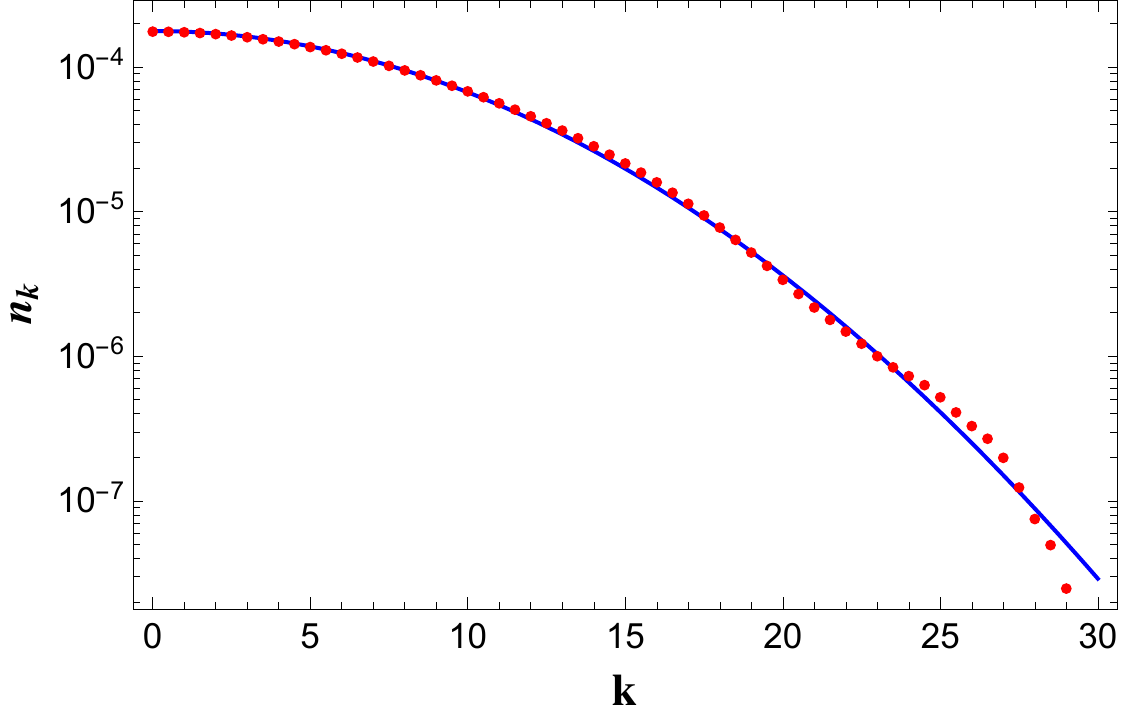}
\caption{Momentum dependence of the produced particle number. The blue solid curve is given by our approximate formula~\eqref{Fpre} with $n=1$. The red dots are the numerical evaluation of the final particle number. We take the parameter set~\eqref{para} and the final time $t_f=30$ in unit of $M^{-1}$. As expected, the final particle number at $t=30$ is the same as that evaluated just after first Stokes line crossing $n=1$. Our analytic estimate is in good agreement with the numerical result.}\label{fig;nk}
\end{figure}

In this case, we are able to evaluate the produced particle number directly from \eqref{Fpre} at $n=1$ since only the first Stokes line crossing around $t\sim \pi$ is responsible for particle production.\footnote{For practical purpose, one may use \eqref{preapp} to estimate the particle number. Here, we will evaluate $n_k\sim e^{-2F_{k,1}^{(0)}}$ using \eqref{Fpre}, which gives slightly better estimate. } We show the comparison between the particle number with our analytic formula~\eqref{Fpre} and the numerical result evaluated at $t=30M^{-1}$ in Fig.~\ref{fig;nk}. The agreement between numerical results and our approximate formula ensures that our method can be reliably applied to obtain the analytic formula for $\beta_k(t)$. 

\begin{figure}[bhtp]
\centering
\includegraphics[width=.6\textwidth]{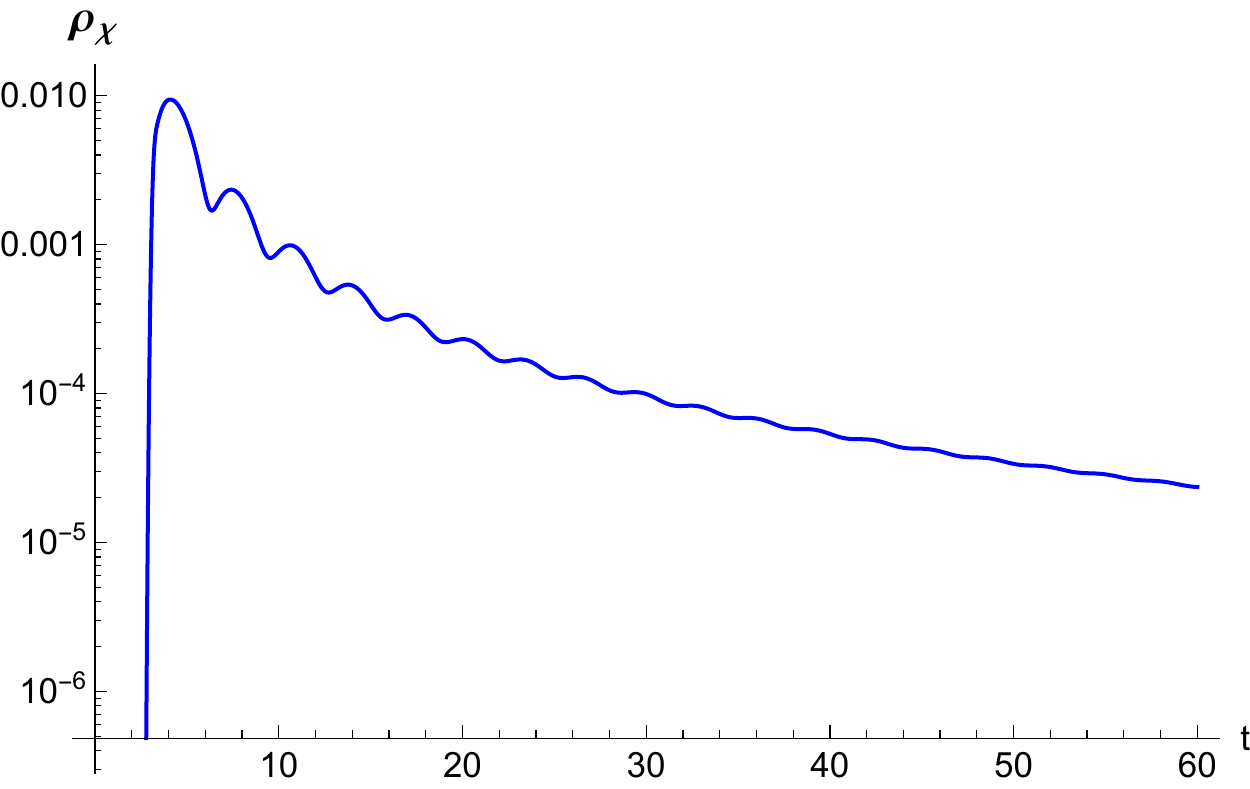}
\caption{Energy density of $\chi$~\eqref{heavyr} with the parameter set~\eqref{para}. The time $t$ is in unit of $M^{-1}$. Note that the oscillatory behavior of $\rho_\chi$ is originated from the oscillation of the effective mass rather than particle production.}\label{fig;rho}
\end{figure}
Finally let us illustrate the approximation quoted in the end of Sec.~\ref{comments}. We consider the time-dependent energy density of $\chi$ with the superheavy mass as the simplest example. Neglecting higher-order terms in adiabatic expansion and in $V_k$, the energy density of $\chi$ is simply given by
\begin{equation}
    \langle\rho_\chi\rangle_{\rm np}\sim \frac{1}{a^3}\int\frac{d^3k}{(2\pi)^3}\omega_k|\beta_k(t)|^2.
\end{equation}
Since the distribution of $\chi$ is centered at $k=0$, we take $k=0$ for the error function in $\beta_k(t)$. Then, using the approximate expression~\eqref{preapp} with $n=1$, we can perform the integration analytically, and find
\begin{equation}
     \langle\rho_\chi\rangle_{\rm np}\sim \frac{\sqrt{\lambda}A(\pi+1)^{4/3}}{8\pi^4 a^2(t)}e^{-\frac{\pi^2m^2}{\sqrt{\lambda}AM}}\left(m^2+\frac{\lambda A^2}{t^2}\sin^2Mt\right)e^{g(t)}K_1(g(t))s(t),\label{heavyr}
\end{equation}
where $K_\nu(x)$ denotes the modified Bessel function of the second kind, and
\begin{equation}
  g(t)\equiv \frac{\pi^2 a^2(t)}{2\sqrt{\lambda}A(\pi+1)^{4/3}}\left(m^2+\frac{\lambda A^2}{t^2}\sin^2Mt\right),  
\end{equation}
\begin{equation}
    s(t)\sim {\rm Erfc}\left(-2m\left(2F_{0,1}^{(0)}\right)^{-1/2}(t-\pi)\right).
\end{equation}
Note that $a(t)=(1+Mt)^{2/3}$ and $F_{0,1}^{(0)}$ in this case is given in~\eqref{Fpre} with $k=0,n=1$. Thus, we could find an approximate expression of the superheavy particle $\chi$ corresponding to the parameter set~\eqref{para}. We show the behavior of the energy density in Fig.~\ref{fig;rho}. Since there is only one event in this case, the evaluation is easily done. In principle, one can perform similar analysis in the models with multiple particle production events either analytically or semi-analytically.

\section{Conclusion}\label{concl}
In this work, we have discussed the superadiabatic basis within cosmological models. As we have reviewed in Sec.~\ref{adrev}, the definition of ``particles'' in intermediate time intervals are quite ambiguous. Nevertheless, there is the optimal choice for describing adiabatic ``particles'' of which particle number changes only near the Stokes line crossing, namely, particle production events. As we show in Sec.~\ref{adpart}, the particle number of the optimal adiabatic basis can be well approximated by our approximate connection formula~\eqref{conmat}, which is an conjectured extension of Dingle-Berry's error function formula~\eqref{DBD}. The superadiabatic basis would also be useful to numerically evaluate the particle number at a intermediate time since non-optimal choices shows badly oscillating values. This point would be clear from our examples. 

We have applied the WKB/phase integral method to various parameter regions of preheating after inflation or its toy model. In particular, by taking Stokes phenomenon into account properly, we could give a unified description for different parameter regions. The narrow resonance seems perturbative effects since the effective particle production rate is analytic in the coupling constant whereas the broad resonance regime shows the production rate to be non-perturbative in the coupling constant. Despite such a different parameter dependence, our approximate formula well describes both regime in a unified manner.

In the resonance regime, where particle production is quite efficient, the interference effects are responsible, and small errors could be amplified in later time as we have discussed. In particular, for the preheating in expanding Universe, it is difficult to obtain the correct interference patterns due to the complicated time-dependence of the effective frequency of particles. Therefore, it seems better to treat the phase factor associated with various Stokes line crossing as random variables. Nevertheless, our method captures the behavior of explosive particle production (see appendix~\ref{stoch}). It would be useful to find the method giving more precise interference patterns as well as to improve our approximate formula~\eqref{conmat}. For this purpose, it would be important to develop the method for the analysis of Stokes phenomenon. 

Despite the above issues, we have found that our approximate formula~\eqref{conmat} can describe the time-dependent particle number consistent with numerical solutions in various parameter regions. In particular, it is important that the approximate solution for $(\alpha_k(t),\beta_k(t))$ are very simple form that enables us to evaluate physical quantities such as energy density of produces particles in analytic or semi-analytic ways. We believe our methods can be applied to various cosmological models. For example, the particle production during inflation may play important roles for inflation dynamics and also leaves some imprints on cosmic microwave backgrounds~\cite{Chung:1999ve,Romano:2008rr,Green:2009ds,Barnaby:2010ke,Cook:2011hg,Flauger:2016idt}. We will leave such applications for future work.

Throughout this work, we have focused on scalar fields, but it is possible to extend this method for particles with different spins. Bosonic particles are straightforward, and the same method is available. For fermions, one is able to apply the superadiabatic basis for two level quantum mechanical system~\cite{Berry_1990,Lim_1991}, which is recently applied to the fermionic particle production in~\cite{Enomoto:2021hfv,Sou:2021juh,Taya:2021dcz}. Since standard model mostly consists of fermions, such an analysis would be necessary for realistic cosmological models.

\acknowledgments
YY would like to thank Soichiro Hashiba for collaboration in our previous work and the early stage of this work, and Minxi He for useful discussions. This work is supported by JSPS KAKENHI, Grant-in-Aid for JSPS Fellows JP19J00494.

\appendix
\section{Universal formula and optimal truncation}\label{univform}
Here, we summarize the approximate formula describing the time evolution of $\beta_k(t)$ near Stokes lines. We do not discuss the derivation of the formula, and refer to Berry's original work~\cite{Barry:1989zz} and recent papers~\cite{Li:2019ves,Sou:2021juh} for derivation of the formula and its application to cosmological models. 

As mentioned in the main text, particle production events can be understood as Stokes phenomena, which lead to the non-trivial mixing of $\alpha_k(t)$ and $\beta_k(t)$ at the Stokes line crossing. Dingle found the universal formula of $\beta_k(t)$ near the Stokes line~\cite{Dingle},
\begin{equation}
    \beta_k(t)\sim \frac{\rm i}{2}{\rm Erfc}(-\sigma_k(t))e^{-F_k^{(0)}},\label{BDf}
\end{equation}
where ${\rm Erfc}(x)$ is the complementary error function, and we have introduced
\begin{eqnarray}
\sigma_k(t)&\equiv& \frac{{\rm Im}F_k(t)}{\sqrt{2{\rm Re}F_k(t)}},\label{sigma}\\
F_k(t)&\equiv&2{\rm i}\int_{t_c}^t\omega_k(t')dt',\label{fkt}\\
F^{(0)}_k&\equiv& {\rm i}\int_{t_c}^{\bar{t}_c}\omega_k(t')dt'.\label{fk0}
\end{eqnarray}
Here $t_c$ denotes the turning point and $\bar{t}_c$ is the conjugate of $t_c$. Note that the phase and branch cut need to be chosen so that $F_{k}^{(0)}$ is positive. After Dingle discovered the error function formula~\eqref{BDf}, Berry showed that the Dingle's universal formula above can be obtained by truncation of the adiabatic expansion, and Borel summation of the remainder terms~\cite{Barry:1989zz}. In Berry's derivation, it is also shown that the optimal order of the adiabatic expansion is given by
\begin{eqnarray}
j\sim {\rm Int}\left[\frac12(F_k^{(0)}- 1)\right],\label{jest}
\end{eqnarray}
where ${\rm Int}[x]$ denotes the Gauss symbol. The comparison of the universal formula with the adiabatic solution at the optimal order is discussed in \cite{Lim_1991} for two level quantum mechanical system with time-dependent Hamiltonian and in \cite{Dabrowski:2014ica,Dabrowski:2016tsx} for strong field QED. 

The above expression is for a single Stokes line crossing, namely a single particle production event. For multiple Stokes line crossings, which would take place in realistic models, such as preheating, we need a generalized formula for the universal approximation. The following generalized formula is proposed in~\cite{Dabrowski:2014ica},
\begin{equation}
    \beta_k(t)=\frac{\rm i}{2}\sum_{n}e^{2{\rm i}\theta_{k,n}}e^{-F_{k,n}^{(0)}}{\rm Erfc}(-\sigma_{k,n}(t)),\label{DBD}
\end{equation}
where 
\begin{eqnarray}
\sigma_{k,n}(t)&\equiv& \frac{{\rm Im}F_{k,n}(t)}{\sqrt{2{\rm Re}F_{k,n}(t)}},\\
F_{k,n}(t)&\equiv&2{\rm i}\int_{t_n}^t\omega_k(t')dt',\\
F^{(0)}_{k,n}&\equiv& {\rm i}\int_{t_n}^{\bar{t}_n}\omega_k(t')dt'.
\end{eqnarray}
Here, $t_n$ and $\bar{t}_n$ are the $n$-th turning point and its conjugate. Note that we have taken the convention that $F_{k,n}^{(0)}>0$. The phase factor $\theta_{k,n}$ is defined as
\begin{equation}
    \theta_{k,n}=\int_{t_i}^{s_n}\omega_kdt, 
\end{equation}
where $t_i$ is the initial time and $s_n$ is the point at which the $n$-th Stokes line crosses the real axis.\footnote{In the original formula proposed in \cite{Dabrowski:2014ica}, the lower limit of the integration in $\theta_{k,n}$ is taken to be the first turning point, but changing it to $t_i$ does not change the relative phase. So, we take the lower limit to be $t_i$ here. } Practically, we may approximate $\theta_{k,n}$ as
\begin{equation}
    \theta_{k,n}\sim \int_{t_i}^{{\rm Re}\ t_n}\omega_k dt,
\end{equation}
where $t_n$ is the $n$-th turning points.

Although the formula~\eqref{DBD} captures the time-dependent particle number correctly for $n_k<1$, we find that it cannot be applied to the case where $n_k>1$ which can be realized by repeated Stokes phenomena such as narrow or broad resonance in preheating. Therefore, we need to improve the approximate formula for $(\alpha_k(t),\beta_k(t))$. Taking the Dingle-Berry formula into account, we propose the following new formula,
\begin{equation}
  \left(  \begin{array}{c} \alpha_k^{(n+1)}\\ \beta_k^{(n+1)}\end{array}\right)=\left(\begin{array}{cc}\sqrt{1+|E_{k,n}(t)|^2}& -{\rm i}E_{k,n}(t)\\ {\rm i}(E_{k,n}(t))^*& \sqrt{1+|E_{k,n}(t)|^2}\end{array}\right)\left(\begin{array}{c} \alpha_k^{(n)}\\ \beta_k^{(n)}\end{array}\right),\label{conmat}
\end{equation}
where $(\alpha_k^{(n)},\beta_k^{(n)})$ denote the Bogoliubov coefficient functions that take into account Stokes line crossing up to the $n$-th event, and 
\begin{equation}
    E_{k,n}=\frac12 e^{-2{\rm i}\theta_{k,n}}e^{-F_{k,n}^{(0)}}{\rm Erfc}(-\sigma_{k,n}(t)).\label{Ek}
\end{equation}
We find this formula a good approximation in several parameter regions including resonance cases, where particle number grows exponentially. Unfortunately, we have not found the derivation of the above formula yet. Nevertheless, let us comment on the reason why the above formula would work in various situations: In certain limit, one finds that this formula corresponds to the connection matrix for the case of parabolic cylinder functions, which has two turning points that are complex conjugate to each other. Since we are mostly considering real potential where there are such pairs of turning points. If there are multiple turning point pairs, each crossing would lead to the mixing of $(\alpha_k(t),\beta_k(t))$. These considerations lead us to the conjectured formula~\eqref{conmat}.

One should iteratively multiply the connection matrix in order to get the Bogoliubov coefficients that take into account up to $n$-th particle production events. We will use this conjectured formula in the main texts and will show that this approximate formula works well.

We note that the formula~\eqref{DBD} can be understood as the replacement of the connection matrix in~\eqref{conmat} as
\begin{equation}
    \left(  \begin{array}{c} \alpha_k^{(n+1)}\\ \beta_k^{(n+1)}\end{array}\right)=\left(\begin{array}{cc}1& 0\\ {\rm i}(E_{k,n}(t))^*& 1\end{array}\right)\left(\begin{array}{c} \alpha_k^{(n)}\\ \beta_k^{(n)}\end{array}\right).
\end{equation}
One can immediately reproduce~\eqref{DBD} from this formula. This approximation means that we neglect the higher-order corrections to $\alpha_k(t)$, which is valid if each $F_{k,n}^{(0)}$ is sufficiently large, or if interference effects does not allow the monotonic growth of $|\beta_k(t)|$. In the case of resonant particle production, one needs to use the form~\eqref{conmat} instead of \eqref{DBD}. We should note however that \eqref{DBD} and \eqref{conmat} may fail to reproduce the correct behavior especially when $F^{(0)}_{k,n}$ are not large enough. We also note that for \eqref{DBD} the normalization condition $|\alpha_k|^2-|\beta_k|^2=1$ cannot be satisfied since the correction to $|\alpha_k|$ is neglected, where as our formula~\eqref{conmat} does. The approximate formula works well for small numbers of particle production events, but small discrepancy with numerical results can occur. In the case of broad resonance regime in an oscillating scalar model, where the value of $F^{(0)}_{k,n}$ is quite small, the small discrepancy between our formula and the numerical result for one event is exponentially increased after multiple events. It would be necessary to improve the accuracy of the formula, but there are possible difficulties: As clearly discussed in~\cite{Taya:2020dco}, the connection problem within (exact) WKB analysis becomes quite difficult in the presence of a real potential term due to the appearance of Stokes segments, which are Stokes lines connecting two turning points. Then, one needs to neglect higher-order instanton contributions. Such higher order terms become relevant when ``instanton action'' $F_{k,n}^{(0)}$ is small. Nevertheless, our formula reproduces numerical results in cases with sufficiently large $F^{(0)}_{k,n}$. Even in the parameter region where the adiabatic expansion seems not good, our formula captures the behavior of time-dependent particle numbers at least qualitatively. Therefore, our connection matrix formula~\eqref{conmat} would be useful to understand the behavior of particle number in various backgrounds.

\section{Emarkov-Milne equation}\label{EMeq}
Following \cite{Dabrowski:2016tsx}, we review the Emarkov-Milne equation, which is an alternative form of the mode equation~\eqref{modeeq}. As an exact solution to~\eqref{modeeq}, we assume the following ansatz
\begin{equation}
    v_k=\xi_k(t)e^{-{\rm i}\lambda_k(t)},
\end{equation}
where $\xi_k(t)$ and $\lambda_k(t)$ are real functions of $t$. The normalization condition $v_k \dot{v}^*_k-\dot{v}_k v^*_k={\rm i}$ reads
\begin{equation}
    \lambda_k=\frac{1}{2}\int_{t_0}^t dt'\xi_k^{-2}(t'),
\end{equation}
where $t_0$ denotes a reference time.
Then, $\xi_k$ should satisfy 
\begin{equation}
    \ddot{\xi}_k(t)+\omega_k^2\xi_k(t)=\frac{1}{4}\xi_k^{-3}.
\end{equation}
Since 
\begin{eqnarray}
    \alpha_k(t)&=&{\rm i}f^{-(j)}_k\left(\dot{v}_k-({\rm i}W^{(j)}_k+V_k)v_k\right),\\
    \beta_k(t)&=&{\rm i}f^{+(j)}_k\left(\dot{v}_k-(-{\rm i}W^{(j)}_k+V_k)v_k\right),
\end{eqnarray}
we are able to rewrite $\alpha_k(t)$ and $\beta_k(t)$ in terms of the amplitude function $\xi_k(t)$ as
\begin{eqnarray}
    \alpha_k(t)=&&\frac{\xi_k}{\sqrt{2W_k^{(j)}}}\left[\frac{1}{2\xi_k^2}+W_k^{(j)}+{\rm i}\left(\frac{\dot\xi_k}{\xi_k}-V_k\right)\right]\nonumber\\
    &&\times \exp\left(-{\rm i}\int_{t_0}^tdt'(2\xi_k^{-2}-W_k^{(j)})\right)\\
    \beta_k(t)=&&-\frac{\xi_k}{\sqrt{2W_k^{(j)}}}\left[\frac{1}{2\xi_k^2}-W_k^{(j)}+{\rm i}\left(\frac{\dot\xi_k}{\xi_k}-V_k\right)\right]\nonumber\\
    &&\times \exp\left(-{\rm i}\int_{t_0}^tdt'(2\xi_k^{-2}+W_k^{(j)})\right).
\end{eqnarray}
For particle number, the phase factor is not relevant and
\begin{equation}
    n_k=\frac{\xi_k^2}{2W_k^{(j)}}\left[\left(\frac{1}{2\xi_k^2}-W_k^{(j)}\right)^2+\left(\frac{\dot\xi_k}{\xi_k}-V_k\right)^2\right]
\end{equation}
This representation is practically useful to numerically obtain the time-dependent particle number for various adiabatic orders as well as for different choices of $V_k$ since we only need to solve $\xi_k$ once. The initial conditions for $\xi_k(t)$ are
\begin{equation}
    \xi_k(t_{\rm ini})\to\frac{1}{\sqrt{2\omega_k(t_{\rm ini})}},\quad \dot{\xi}_k(t_{\rm ini})\to0.
\end{equation}
We note that if $\omega_k(t)$ is taken to be asymptotically time-independent as $t\to t_{\rm ini}$, the initial particle number (or equivalently $\beta_k(t_{\rm ini})$) becomes zero for any adiabatic expansion order since all the higher-order terms vanish. However, an oscillating scalar background, for example, does not have such behavior. Therefore, one finds non-zero initial particle number, which is unphysical. Therefore, one should subtract such contribution, which we performed in all the numerical simulations shown in this work.

\section{Perturbative decay of an oscillating scalar field}\label{pertWKB}
We discuss the perturbative decay from scalar condensate of $\phi$ to a massive spectator scalar $\chi$ in the presence of the three point coupling $\mathcal{L}_{\rm int}=\frac12 \lambda \phi \chi^2$. We assume that $\phi(t)=A \cos Mt$ where $M$ being a mass of $\phi$ and $A$ a constant corresponding to the amplitude. We take the background spacetime to be Minkowski spacetime, and consider a perturbation theory with respect to $\lambda$. Our setup corresponds to $h(\phi)=A\cos(Mt)$ and $g_{\mu\nu}=\eta_{\mu \nu}$ in \eqref{action}. We assume $2m<M$ so that the decay of $\phi$ into two $\chi$ is kinematically allowed. Here, we take WKB choice, namely $W_k^{(j)}=\omega_k$ and $V_k=0$, where the effective frequency $\omega_k$ is given by
\begin{equation}
    \omega_k^2=k^2+m^2+\lambda\phi(t).
\end{equation}
In the following we assume $|\lambda A|<k^2+m^2$ such that $\omega_k^2$ does not experience instability. 

At the first order in $\lambda$, the effective frequency can be expanded as
\begin{equation}
    \omega_k\sim \Omega_k+\frac{\lambda \phi}{2\Omega_k},
\end{equation}
where $\Omega_k=\sqrt{k^2+m^2}$. With this expansion, the leading order equation for Bogoliubov coefficient functions $\alpha_k(t)$ and $\beta_k(t)$ are found to be
\begin{equation}
   \dot{\alpha}_k\sim \frac{\lambda\dot\phi}{4\omega_k^2}ce^{2{\rm i}\Omega_k t},\qquad \dot{\beta}_k\sim \frac{\lambda \dot\phi}{4\Omega_k^2}\bar{c}e^{-2{\rm i}\Omega_k t},
\end{equation}
where $c=e^{2{\rm i}\Omega_k t_0}$ and $t_0$ is the initial time. Taking initial conditions $\alpha_k(t_0)=1$ and $\beta_k(t_0)=0$, we find the leading order solutions of the above differential equation as
\begin{equation}
\alpha_k(t)\sim 1, \qquad \beta_k(t)\sim \int_{t_0}^t dt'\frac{\lambda \dot{\phi}(t')}{4\Omega_k^2}\bar{c}e^{-2{\rm i}\Omega_k t'},
\end{equation}
and these solutions satisfy $|\alpha_k|^2-|\beta_k|^2=1+{\mathcal O}(\lambda^2)$.
Thus, we find a perturbed positive frequency solution to be
\begin{eqnarray}
    f_k&=&\frac{\alpha_k}{\sqrt{\omega_k}} e^{-{\rm i}\int\omega_k dt}+\frac{\beta_k}{\sqrt{\omega_k}} e^{{\rm i}\int\omega_k dt}\nonumber\\
    &\sim&\frac{\bar{c}^{1/2}}{\sqrt{2\Omega_k}}e^{-{\rm i}\Omega_kt}\left(1-\frac{\lambda\phi}{4\Omega_k^2}-{\rm i}\int_{t_0}^t\frac{\lambda\phi(t')dt'}{2\Omega_k}\right)\nonumber\\
    &&+\frac{\bar{c}^{1/2}}{\sqrt{2\Omega_k}}e^{{\rm i}\Omega_kt}\int_{t_0}^t\frac{\lambda\dot{\phi}(t')dt'}{2\Omega_k}e^{-2{\rm i}\Omega_kt'}.
\end{eqnarray}
One can easily check that this perturbed solution satisfies the mode equation up to $\mathcal{O}(\lambda^2)$.

The particle production rate can be found from the perturbed $\beta_k$ as follows: We consider particle number produced during the time interval sufficiently longer than $\phi$-oscillation periodicity $\sim M^{-1}$. Then, we may take the integration limit to be infinite, and find
\begin{equation}
    \beta_k\sim \frac{{\rm i}\lambda\bar{c}AM\pi}{4\Omega_k^2}\delta(M-2\Omega_k),
\end{equation}
where we have used $\dot\phi=-AM\sin Mt$ and dropped another term proportional to $\delta(M+2\Omega_k)$. Therefore, 
\begin{equation}
    |\beta_k|^2=\frac{\lambda^2A^2M^2\pi^2}{16\Omega_k^4}(\delta(M-2\Omega_k))^2.
\end{equation}
The square of the delta-function should be understood as
\begin{equation}
   \delta(M-2\Omega_k)\delta(0)=\frac{T_{\rm tot}}{2\pi}\delta(M-2\Omega_k),
\end{equation}
where $T_{\rm tot}$ ($MT_{\rm tot}\gg1$) denotes the total time interval, which is taken to be infinity in evaluating the time integral. 
We can easily perform the integration over momentum space as
\begin{eqnarray}
    n_{\rm tot}&=&\int \frac{d^3k}{(2\pi)^3}|\beta_k|^2\nonumber\\
   &=& \frac{\lambda^2A^2}{32\pi}\sqrt{1-\frac{2m^2}{M^2}}T_{\rm tot},
\end{eqnarray}
where $n_{\rm tot}$ denotes the particle density per unit volume. Therefore, the $\chi$-particle production rate per unit time~$\Gamma_\chi$ is
\begin{equation}
    \Gamma_\chi=\frac{\lambda^2A^2}{32\pi}\sqrt{1-\frac{2m^2}{M^2}}.
\end{equation}
A perturbative decay rate of $\phi$ to two $\chi$ is given by \begin{equation}
    \Gamma_{\phi\to \chi\chi}=\frac{\lambda^2}{32\pi M}\sqrt{1-\frac{2m^2}{M^2}},
\end{equation}
and therefore, we may rewrite the production rate $\Gamma_\chi$ as
\begin{equation}
    \Gamma_\chi=2n_\phi\Gamma_{\phi\to \chi\chi},
\end{equation}
where $n_\phi=\frac{1}{2}MA^2$ is the number density of zero-mode $\phi$-particle.
This is consistent with the small amplitude limit results in \cite{Yoshimura:1995gc,Matsumoto:2007rd}, where the production of $\chi$-particles from coherent state scalar condensate was considered. Thus, we can find perturbative decay rate of $\phi$-condensate to $\chi$-particles with the use of WKB solutions. This result is reasonable since our perturbative calculation corresponds to the insertion of a zero mode of $\phi$ to $\chi$'s propagator / mode function, which leads to the nontrivial time-dependence of $\beta_k(t)$.

We should note that this evaluation of particle production is quite different from neither broad nor narrow resonance that can be described by Stokes phenomenon (see Sec.~\ref{scalarosc}).
\begin{figure}[htbp]
\centering
\includegraphics[width=.7\textwidth]{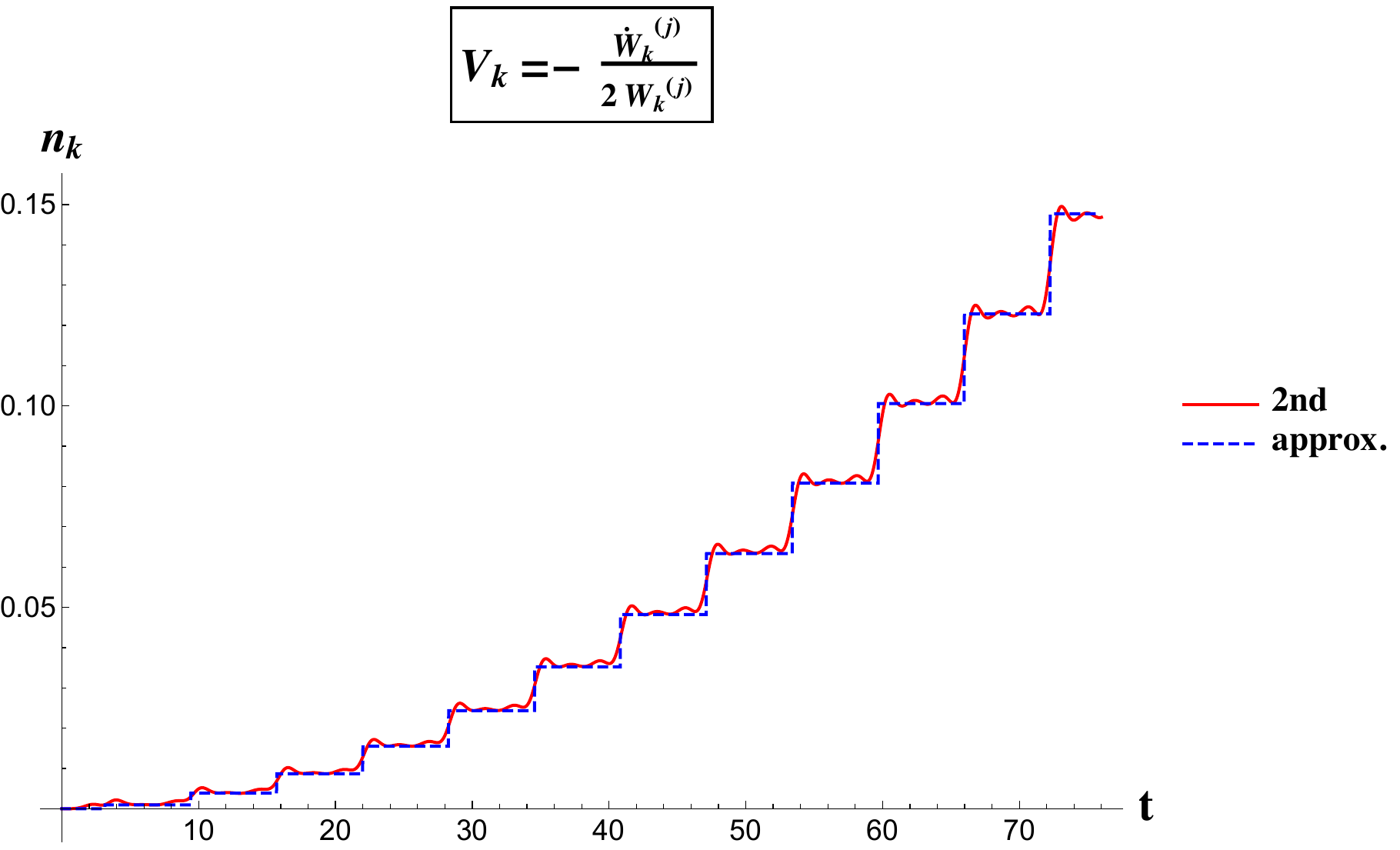}
\caption{Time-dependent particle number for the narrow resonance case with a modified approximate formula. We have taken $k^2+m^2=1/3.99, M=1, \lambda A= 0.01$, and the adiabatic order is the second order with the natural choice of $V_k$. We should note that we put the overall factor $1/1.4$ to the approximate formula.}\label{fig;narrow2}
\end{figure}
\section{Narrow resonance with different approximation}\label{narrowmod}
In the narrow resonance case with the second order adiabatic expansion, we find step function like behavior in growth of particle number. Let us examine a replacement of error function formula by Heaviside $\Theta$ function, which actually gives a better fitting to the numerical result. More explicitly, we replace the function $E_{k,n}(t)$ in \eqref{resapp} with
\begin{equation}
   \tilde{E}_{k,n}=e^{2{\rm i}n\theta}e^{-F_k(0)} \Theta(t-(2n-1)\pi),
\end{equation}
where $\Theta(x)$ is the Heaviside $\Theta$ function. Then, we find the behavior shown in Fig.~\ref{fig;narrow2}. The modified formula is not derived from partial Borel resummation performed by Berry. However, we expect that the exact WKB analysis would lead to a similar formula since in the exact WKB analysis, the adiabatic series is not Borel summable precisely on the Stokes line, which would look like a boundary of regions in complex time plane.

\section{Stochastic resonance in expanding Universe}\label{stoch}
\begin{figure}[htbp]
\centering
\includegraphics[width=.7\textwidth]{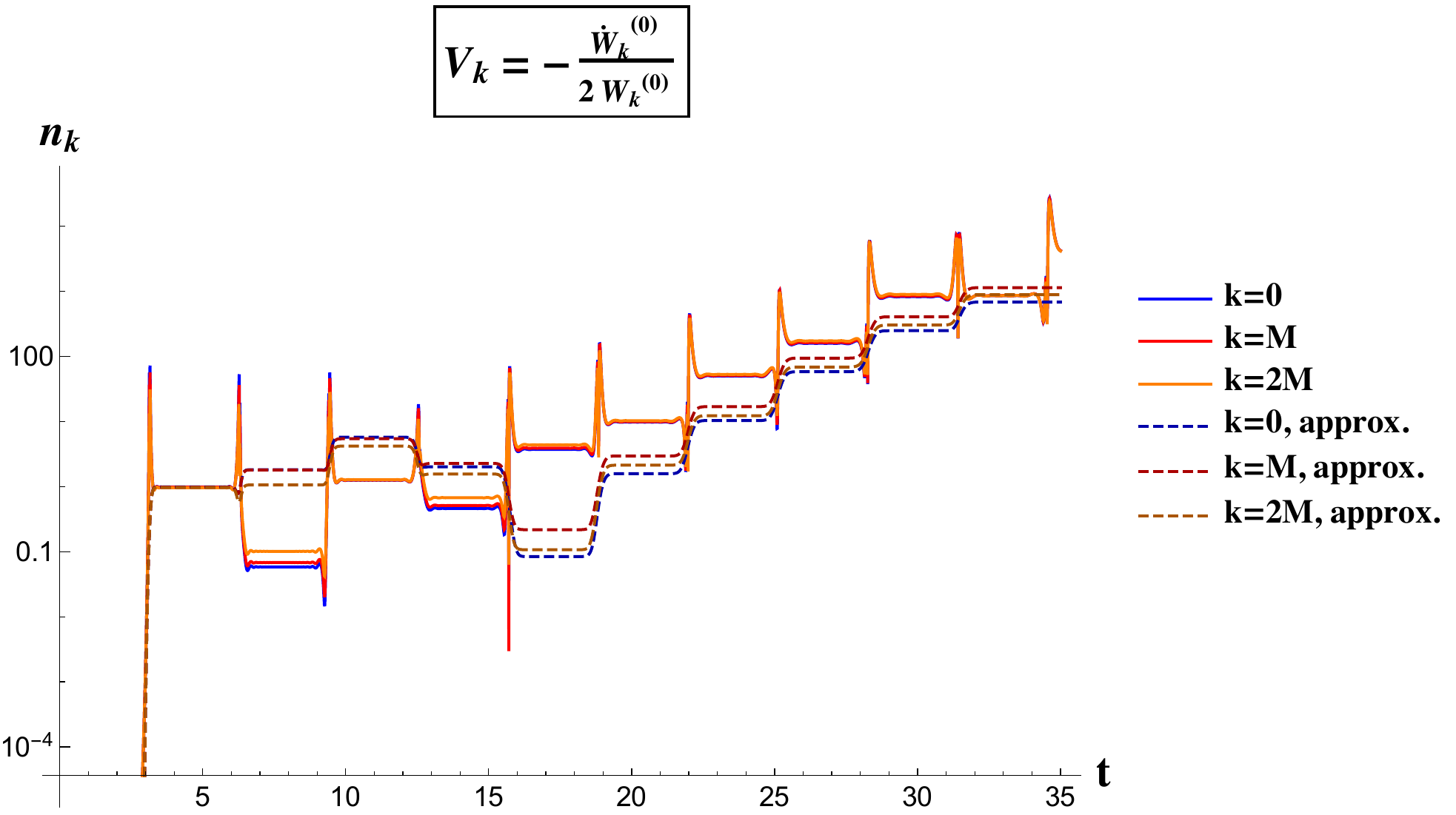}
\caption{Time-dependent particle number in stochastic resonance regime in the expanding Universe. The solid lines are numerical results and the dashed lines are analytic approximation based on \eqref{epre} and \eqref{conmat}. We used $A=650$, $c=-0.1$, $m=0.5M$, and $k=0,M,2M$. The time $t$ is in unit of $M^{-1}$.}\label{fig;preheat}
\end{figure}
In this section, we examine whether our analytic formula~\eqref{Fpre} captures the behavior of particle number in stochastic resonance in expanding Universe discussed in Sec.~\ref{preheating}. As we explained in the main text, it is quite difficult to reproduce numerical solutions in this case. Unlike the case of $F_{k,n}^{(0)}$, it is difficult to make a good approximation for the phase factor $\theta_{k,n}\sim\int_{t_i}^{t_n}\omega_kdt$ where $t_{i,n}$ denote the initial time and the real part of $n$-th turning point, respectively. In addition to this issue, the connection formula~\eqref{conmat} does not work well when $F_{k,n}^{(0)}$ due to the higher order instanton corrections. Nevertheless, we would like to discuss whether our analytic approach captures the stochastic resonance behavior. For phase factors, we will take
\begin{equation}
    \theta_{k,n}=\int_{0}^{n\pi M^{-1}+c}\sqrt{\frac{A^2\sin^2 t}{t^2}}+n\pi\sqrt{m^2+\frac{k^2}{(1+n\pi)^{4/3}}},\label{thapp}
\end{equation}
where $c$ is a constant parameter. This is not an approximation of the original integration, but a formula to be taken for illustrative purposes. The function $E_{k,n}(t)$ in this case is given by
\begin{equation}
    E_{k,n}(t)=\frac{1}{2} e^{-F^{(0)}_{k,n}-2{\rm i}\theta_{k,n}}{\rm Erfc}(-\sigma_{k,n}(t)),\label{epre}
\end{equation}
where $F_{k,n}^{(0)}$ is that in \eqref{Fpre}, and $\sigma_{k,n}$ is approximately given by
\begin{equation}
    \sigma_{k,n}(t)\sim \frac{2\sqrt{m^2+\frac{k^2}{(1+n\pi)^{4/3}}}}{\sqrt{2F_{k,n}^{(0)}}}(t-n\pi).
\end{equation}
Then, one can obtain the time-dependent Bogoliubov coefficients by~\eqref{conmat}. Using the derived $\beta_k(t)$, we show how the approximate particle number behaves in Fig.~\ref{fig;preheat}. As seen in the figure, although the value shows large discrepancy between analytic and numerical results, the transition behavior looks similar. Our analytic approximation captures the particle number increasing or decreasing in time, but eventually amplified to large number. It fails to reproduce the numerical result, but at the first event $t\sim \pi$, all the lines coincide. This means that the particle number estimation based on \eqref{Fpre} is correct but interference effects are not correctly reproduced. We should note that in Fig.~\ref{fig;preheat} although the final values of particle number of the approximation and numerical ones seem to coincide, this is just a coincidence.


\bibliographystyle{JHEP}
\bibliography{ref.bib}

\end{document}